\documentclass[preprint,5p,twocolumn]{elsarticle}

\usepackage{amssymb}
\usepackage{amsmath}

\usepackage{lineno}
\usepackage{lineno,hyperref}
\usepackage{booktabs}
\usepackage{graphicx}
\usepackage{multirow}
\usepackage{tabularray}
\usepackage{float} 
\usepackage{makecell}
\modulolinenumbers[1]
\usepackage{tabularx}
\usepackage{pgfplots}
\usepackage{stfloats}
\usepackage{amsmath}
\usepackage{stfloats}
\usepackage{subcaption}
\journal{Remote Sensing of Environment}

\begin{document}

\begin{frontmatter}
\title{Development of an Uncertainty Workflow to Support Landsat TIRS Split Window-Derived Surface Temperature Products} 

\author[rit]{Amirhossein Hassanzadeh\corref{cor}}
\ead{axhcis@rit.edu}
\author[rit]{Robert Mancini}
\author[rit]{Aaron Gerace}
\author[rit]{Rehman Eon}
\author[rit]{Matthew Montanaro}

\affiliation[rit]{organization={Chester F. Carlson Center for Imaging Science, Rochester Institute of Technology},
            addressline={}, 
            city={Rochester},
            postcode={14623}, 
            state={NY},
            country={USA}}

\cortext[cor]{Corresponding author}

\begin{abstract}

   The current Landsat Level~2 surface temperature products are derived using a single-channel (SC) methodology to estimate per-pixel surface temperature (ST) maps from Level~1 radiance data.  The corresponding uncertainty map workflow utilizes standard error propagation in conjunction with a cloud-proximity scheme to derive per-pixel uncertainty estimates associated with the temperature maps. A known issue with the Level~2 uncertainty, however, is its susceptibility to overestimation of uncertainty due to its dependence on Landsat's cloud mask, which is prone to false-positives. These false positives most commonly occur on cloud-free days, which represents a favorable condition for accurately estimating surface temperature from spaceborne thermal instruments. Beginning with Collection 3, the split window (SW) approach will serve as the surface temperature algorithm for the level-2 product, reflecting its adaptability across conditions which necessitates the development of a dedicated uncertainty workflow. In this study, we introduce an improved uncertainty workflow, based on a physical parameter called total precipitable water (TPW), that more adequately estimates the uncertainty associated with surface temperature estimates. Considering this process was developed for the dual-channel Thermal Infrared Sensor (TIRS)~1~\&~2 instruments onboard Landsat~8~\&~9, we leveraged a SW algorithm for estimating surface temperature to drive the uncertainty methodology discussed here. First, considering Landsat is not equipped with the optical channels necessary for deriving TPW, we create an XGBoost-based machine learning pipeline that relates TIRS bands 10 \& 11 image radiance to TPW using the MODIS product as reference. We find that performing robust feature engineering and feature selection significantly improves algorithm performance within balanced datasets. The resulting modeling approach achieves a mean absolute error in estimating TPW of 0.54 [cm] and a coefficient of determination (R²) as high as 0.89. Secondly, we propose an improved (SW-based) uncertainty workflow that also uses standard error propagation but incorporates uncertainty as a function of TPW. Validation of this workflow shows that the proposed algorithm better represents the uncertainty associated with ST retrieval than the existing Level~2 workflow. We provide the derived coefficients that are required to apply SW and the corresponding uncertainty workflow. Our work fills the gap in the operational surface temperature algorithms and their corresponding uncertainty workflow tailored for Landsat 8 and 9, and machine learning based models for predicting atmospheric water vapor using thermal infrared sensor bands on board Landsat 8 and 9.  

\end{abstract}

\begin{highlights}
\item Landsat Collection 3 will use the Split Window algorithm for Level-2 ST.  
\item Current Landsat Level-2 SC uncertainty product is prone to false positives.  
\item New uncertainty workflow for Split Window integrates total precipitable water.  
\item Machine learning estimates TPW from Landsat TIRS bands 10 and 11 accurately. 
\item Results show improved reliability of uncertainty in surface temperature retrieval.
\end{highlights}


\end{frontmatter}

\section{Introduction}

\begin{figure*}
\centering
\includegraphics[width=1.0\linewidth]{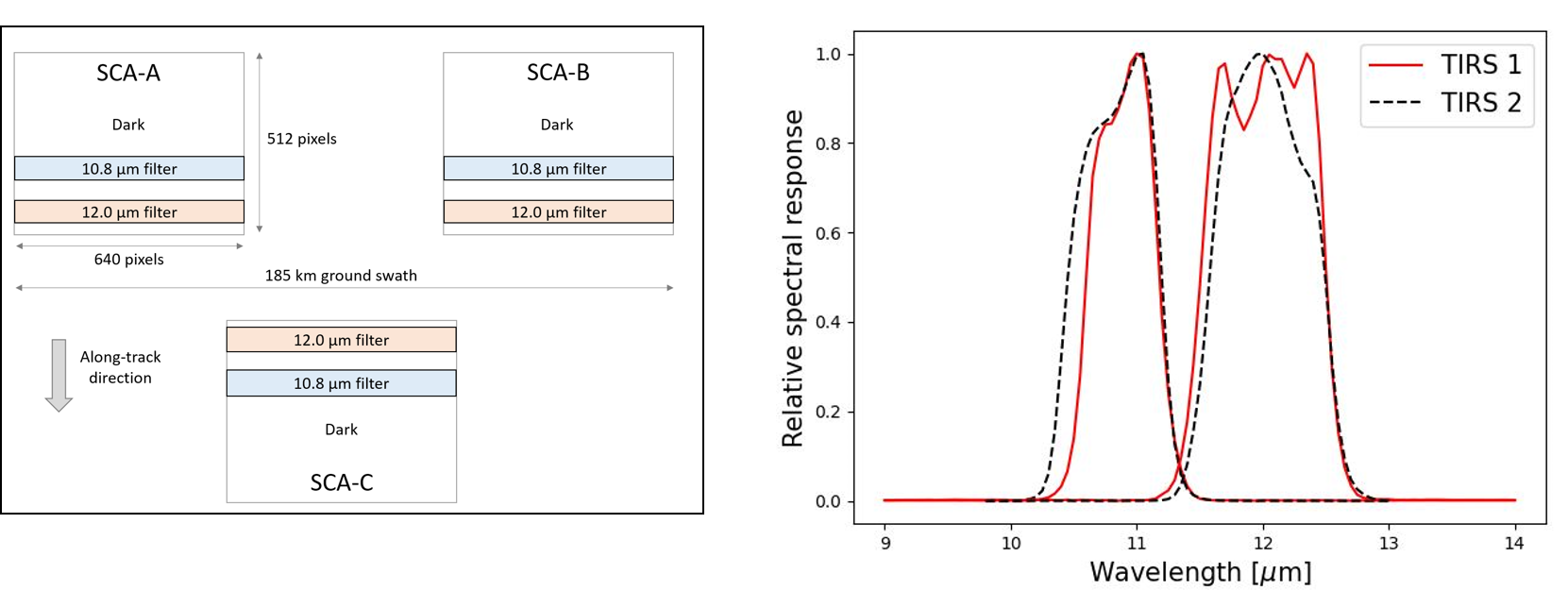}
\caption{(left) Illustration of the TIRS-class focal plane assembly, (right) relative spectral response functions of TIRS 1 and TIRS 2.}
\label{fig:intro:rsr}
\end{figure*}

The launch of Landsat 9 (L9) in 2021 marked another chapter in the Landsat program, extending its already impressive archive. 
The sensors onboard L9 are functional copies of those onboard Landsat 8 (L8), except for baffles that were integrated into the Thermal Infrared Sensor (TIRS)-2 to mitigate any effects of stray light, observed in the TIRS-1 imagery data~\cite{montanaro2014stray}. As such, the algorithms developed for L8 image data can be applied to L9 with minor modifications to account for the differences in the instruments’ spectral responses. An illustration of the focal-plane assembly (left) and the Relative Spectral Response (RSR) functions (right) for the TIRS-class sensors are shown in Figure \ref{fig:intro:rsr}. 

Among the available Landsat products in Collection~2 is the Level~2 Surface Temperature (ST) product, which currently employs a single-channel (SC) algorithm to derive surface temperature for the Landsat archive~\cite{laraby2018uncertainty, laraby2017landsat}. However, the SC approach is susceptible to false positives in the associated uncertainty maps~\cite{laraby2018uncertainty}. On the other hand, the split-window (SW) approach, widely used in ST retrievals, offers improved accuracy by leveraging the differential absorption characteristics between two thermal bands~\cite{mcmillin1975estimation,galve2008atmospheric}. Given its robustness, starting with Collection 3, the SW approach will be adopted as the surface temperature algorithm for the level-2 product. Consequently, it requires the development of a dedicated uncertainty workflow.

Proper error propagation through the model’s various terms enables quantification of the SW algorithm’s uncertainty. To further improve the reliability of uncertainty maps, an enhanced uncertainty estimation framework can be implemented—integrating atmospheric water vapor information which has shown to improve the accuracy of SW uncertainty~\cite{galve2008atmospheric}.

In this study, our primary objective is to explore the uncertainty components of the SW algorithm in greater depth, analyze the contributing terms, and incorporate atmospheric water vapor to improve the accuracy of the uncertainty estimation workflow. We also build upon the previously introduced SW approach~\cite{gerace2020towards} for L8, by reporting updated coefficients for L9. 

\noindent \textbf{Main contributions.} We summarize our research contributions as follows:

\begin{itemize}

    \item \textbf{Enhanced SW uncertainty workflow.} We introduce an improved SW uncertainty workflow for L9 that accounts for various sources of error such as algorithmic error, sensor noise, and surface emissivity—through proper error propagation. Using the Thermodynamic Initial Guess Retrieval (TIGR~\cite{tigr}) atmospheric profile database and Moderate resolution atmospheric Transmission (MODTRAN~\cite{10.1117/12.2050433}) simulations across varying water vapor conditions, we derive a relationship between total precipitable water (TPW) and SW algorithmic error. This relationship enables the use of TPW estimates as input for predicting per-pixel SW algorithmic uncertainty. 

    \item \textbf{Machine learning based TPW estimation.} To obtain the per-pixel TPW estimates required by the uncertainty framework, we develop generalizable a machine learning model that predicts TPW from coincident L9 TIRS-2 data as inputs and Moderate Resolution Imaging Spectrometer (MODIS) TPW product as ground truth. We employ a boosting-based approach (XGBoost~\cite{chen2016xgboost}).

    \item \textbf{Robust TPW data processing and feature optimization pipeline.} We construct a robust modeling pipeline starting with the collection of coincident L9 TIRS-2 and MODIS TPW data, followed by comprehensive masking and filtering to ensure high-quality input/output. We propose a pipeline that integrates mathematical feature engineering and feature selection (using Jostar;~\cite{hassanzadeh2021broadacre}), enabling input dimensionality reduction for improved generalization.

    \item \textbf{Tailored SW approach for L9.} We adapt and tailor the SW methodology for use with the TIRS-2 sensor onboard L9. We derive the SW coefficients for L9 using MODTRAN modeling of different atmospheric profiles using the TIGR database.

\end{itemize}
\subsection{Background}
L8 and L9 data are available in two collection tiers—Collection 1 and Collection 2—each offering two primary product levels: Level 1 and Level 2. Collection 2 features improved radiometric and geometric calibration and updated processing algorithms compared to Collection 1. On the other hand, Level 1 products are corrected for sensor and geometric distortions, while Level 2 products undergo atmospheric correction. Among the Level 2 products are surface reflectance (SR), ST, and vegetation indices (VIs)~\cite{WULDER2022113195, product_usgs}. ST product, specifically, plays an important role in water resource management and crop mapping through evapotranspiration, disaster management, and ecological observations~\cite{ANDERSON201250, deliry2020assessment, guha2018analytical}. 

\begin{figure*}[ht]
    \centering
    \includegraphics[width=1\textwidth]{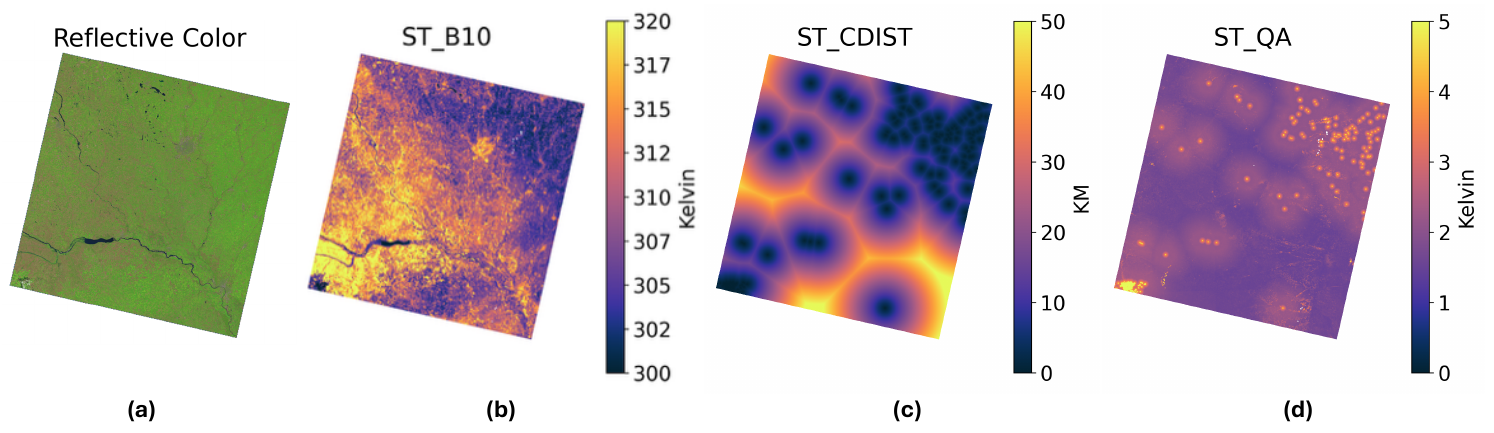}
    \caption{(a) RGB color representation of a sample Landsat 8 scene with no visible clouds. (b) Corresponding Level 2 Surface Temperature (ST) product generated using the SC algorithm. (c) Associated DTC map, highlighting the presence of false positives. (d) SC ST QA band representing uncertainty in the ST retrieval. We can see that despite the absence of clouds in the scene, false positive cloud detections negatively impact the uncertainty map. Scene Product ID:LC08\_L1TP\_029030\_20220721\_20220801\_02\_T1; Cloud Cover = $0.08\%$.}
    \label{fig:dtc_false_positives}
\end{figure*}

The current Level 2 ST product relies on the SC algorithm and a broad-band infrared channel (Band 10) to estimate land surface temperature. The associated SC uncertainty estimation workflow takes advantage of distance to cloud (DTC) metric, also know as surface temperature cloud distance (ST\_CDIST) product. DTC map is derived from the cloud mask generated by C Function of Mask (CFMask;~\cite{foga2017cloud}), with pixel values representing proximity to a cloud; the lower the value, the closer the cloud. Due to dependence of SC uncertainty workflow on DTC, the resulting uncertainty map increases as a pixel’s proximity to a detected cloud decreases (see Figure 6 in~\cite{laraby2018uncertainty}). Furthermore, the CFMask performing at 95\% accuracy~\cite{foga2017cloud} can result in false positives, where highly reflective surfaces with cold apparent temperatures—such as sand or aluminum—may be misclassified as clouds. With respect to the uncertainty workflow and the derived DTC map from cloud mask, this negatively affects the neighboring pixels leading to inaccurate uncertainty estimates (see Figure 16 in~\cite{laraby2018uncertainty}). Figure~\ref{fig:dtc_false_positives} shows the RGB representation of a cloud free L8 scene, the corresponding SC ST product, the ST\_CDIST product, and the associated SC ST uncertainty map. This figure illustrates that a substantial number of pixels are adversely impacted by the cloud mask and the DTC metric, resulting in an unreliable per-pixel uncertainty characterization; pixels falsely identified as clouds propagate their influence to neighboring pixels by assigning them a low distance-to-cloud value, which in turn inflates the SC ST uncertainty map.

The SW algorithm estimates surface temperature using two long-wave infrared bands and has been widely applied across various sensors and domains, including sea surface temperature retrieval, land surface temperature estimation, and atmospheric correction~\cite{becker1990towards, wan1996generalized, coll1994atmospheric, ulivieri1994split}. SW was implemented during early satellite missions such as Advanced Very High Resolution Radiometer (AVHRR)~\cite{llewellyn1984satellite} and MODIS~\cite{brown1999modis}. The SW algorithm has been validated in multiple studies using data from Surface Radiation Budget Network (SURFRAD) sites and buoy-based ST measurements across the United States~\cite{duan2021validation, malakar2018operational}. For Landsat’s TIRS-class sensors, Gerace et al. (2020) proposed an architecture that builds on prior work to enable an operational ST product~\cite{gerace2020towards}. More recently, Eon et al. (2023) conducted an under-flight experiment as part of a cross-validation study between L8 and L9 SW-derived ST and unmanned aerial system (UAS)-based thermal imagery~\cite{eon2023validation}. Their findings report maximum temperature differences of 1 K over water and 2 K over sand. This highly-utilized algorithm has been adapted and modified for a variety of sensors and can be trained for specific applications ~\cite{gerace2020towards,mao2005practical,kerr1992accurate,jiang2008split}. 

In this study, we apply the generalized SW algorithm originally proposed by Wan and Dozier (1996)~\cite{wan} to derive ST from imagery acquired by TIRS sensors, as shown in Equation~\ref{eq:methods:SW-equation1}, where,

\begin{figure*}[b]
\begin{equation}
\begin{aligned}
    ST = & \quad b_0 + \left(b_1+b_2 \frac{1-\epsilon}{\epsilon}+b_3 \frac{\Delta \epsilon}{\epsilon^2}\right) \frac{T_i+T_j}{2} + \left(b_4+b_5 \frac{1-\epsilon}{\epsilon}+b_6 \frac{\Delta \epsilon}{\epsilon^2}\right) \frac{T_i-T_j}{2} +b_7(T_i-T_j)^2,
    \label{eq:methods:SW-equation1}
\end{aligned}
\end{equation}
\end{figure*}

\begin{itemize}
\setlength\itemsep{-0.5em}
\item[--] $ST$ is the desired surface temperature $\left[K\right]$
\item[--] $b_k$ (for k = 0,1,...,7) are sensor and algorithmic coefficients derived through a training process using MODTRAN
\item[--] $i$ and $j$ correspond to the two thermal bands (Bands 10 and 11 for TIRS)
\item[--] $\Delta \epsilon = \epsilon_i-\epsilon_j$ is the difference in band-effective emissivities
\item[--] $\epsilon = (\epsilon_i+\epsilon_j$)/2 is the average of the band-effective emissivities
\item[--] $T_i, T_j$ are the apparent temperatures in the two thermal bands
\end{itemize}

The corresponding uncertainty workflow associated with this SW approach can be quantified through standard error propagation. This process involves calculating the contribution of each term by taking partial derivatives of each term as well as accounting for correlations between terms; as shown in Equation~\ref{equation:sw_uncertainty}. The standard error propagation allows us to incorporate errors from multiple sources, including brightness temperature measurements, surface emissivity, and the uncertainty associated with the fit coefficients ($b_k$), evaluated separately for L8 and L9. We provide a detailed explanation of the SW uncertainty procedure in the methodology section of this study.

\begin{figure*}[t]
\begin{equation}
    \delta ST = \sqrt{
    \begin{aligned} &
    \delta b^2 +  \left(\frac{\partial ST}{\partial T_i}\delta T_i \right)^2 + \left(\frac{\partial ST}{\partial T_j}\delta T_j \right)^2 +
    \left(\frac{\partial ST}{\partial \epsilon_i}\delta \epsilon_i \right)^2 +
    \left(\frac{\partial ST}{\partial \epsilon_j}\delta \epsilon_j \right)^2 \\ & +  
    2\rho(T)_{ij} \frac{\partial ST}{\partial T_i}\frac{\partial ST}{\partial T_j}\delta T_i \delta T_j + 
    2\rho(\epsilon)_{ij} \frac{\partial ST}{\partial \epsilon_i}\frac{dST}{\partial \epsilon_j}\delta \epsilon_i \delta\epsilon_j 
    \label{equation:sw_uncertainty}
    \end{aligned}}
\end{equation}
\end{figure*}

Related research has shown that explicitly accounting for atmospheric water vapor improves the accuracy of the SW algorithm~\cite{wan1996generalized, ottle1997estimation, harris1992extension, ulivieri1994split, yu1994non, franccois2002atmospheric}. Several approaches have leveraged the covariance-to-variance ratio of brightness temperatures to estimate column water vapor, thereby enhancing SW algorithm performance~\cite{ottle1997estimation, sobrino1994improvements}. Ottlé (1997) reported that using covariance-to-variance method resulted in a root mean square error (RMSE) of less than 0.5 g/cm² in estimated total precipitable water (TPW) when validated against data from the Along Track Scanning Radiometer (ATSR) and AVHRR~\cite{ottle1997estimation}. Similarly, Liu et al. (2017) introduced a modification to the generalized SW algorithm by incorporating a third band centered at 6.7 µm, specifically for water vapor estimation~\cite{liu2017improved}. This approach was validated using MODIS and the Stretched Visible and Infrared Spin Scan Radiometer onboard FengYun-2G (SVISSR/FY-2G).

In a related effort, the study by Galve et al. (2008) proposed a modified SW algorithm for MODIS, incorporating surface emissivity, temperature, and atmospheric water vapor as independent variables, with emissivity and temperature terms being decoupled~\cite{galve2008atmospheric}. Their results showed that, for MODIS, the error between estimated ST and ground reference increased with higher atmospheric water vapor content (see Figure 5 in~\cite{galve2008atmospheric}). They conducted a detailed error propagation analysis accounting for uncertainties in brightness temperature, emissivity, water vapor, and algorithmic fitting coefficients, with total error estimates around 0.4 K. For real-time applications, the study utilized MOD07 data to obtain total precipitable water estimates to support operational ST retrievals.

Landsat is not equipped with the optical bands necessary to directly measure TPW. As such, the use of a ratio-based approach for TPW retrieval (similar to 2-channel and 3-channel ration approach for MODIS~\cite{gao2003water}) is not feasible. An accurate retrieval of TPW for L8 and L9 likely requires relying on external products from instruments specifically designed to sense water vapor. MODIS provides the MOD05 product TPW at 1 km ground sampling distance (GSD). Prior to the Constellation Exit Maneuver in October 2022~\cite{modis_exit}, MODIS and L9 had nearly coincident overpasses, with a maximum temporal offset of 30 minutes. This temporal alignment presents an opportunity to apply supervised machine learning methods capable of detecting complex nonlinear patterns in the data and develop a mapping function that leverages TIRS Bands 10 and 11 image data to estimate per-pixel TPW using the MODIS TPW product as truth. 

The proposed approach shifts the dependency of Landsat’s current Level-2 ST uncertainty workflow away from the DTC parameter and instead bases it on per-pixel TPW estimates. The relationship between TPW and algorithmic error can first be established using the MODTRAN radiative transfer model (RTM) and expressed as a function:

\begin{equation}
\text{SW\_ST} - \text{LST} = f(\text{TPW})
\end{equation}

where the algorithmic error represents the difference between the SW estimated ST (SW\_ST) and the true LST, and $f(\cdot)$ is a function derived from MODTRAN simulations. During operational processing, a supervised machine learning model can estimate per-pixel TPW from TIRS image data. The corresponding algorithmic error can then be inferred using the pre-established function $f(\text{TPW})$, and this estimated error can be incorporated into the broader uncertainty propagation workflow. We claim that that this approach preserves the magnitude of the uncertainty and removes the unanticipated artifacts originating from an imperfect cloud mask and the associated DTC metric.

\section{Methods}

In this section, we revisit key methodologies previously published in relation to the SW workflow and present the corresponding coefficients for the TIRS-2 sensor onboard L9~\cite{gerace2020towards}. We then examine the SW uncertainty workflow in detail, outlining the contributing terms using the standard error propagation approach. This includes establishing a link between the SW algorithmic error and TPW, derived from MODTRAN simulations and the TIGR atmospheric database. Finally, we describe a modeling approach for TPW estimation based on the top-of-atmosphere (TOA) radiance of TIRS Bands 10 and 11, using MODIS TPW as a reference. This enables per-pixel TPW estimation, which is subsequently used to characterize the algorithmic error in the SW uncertainty workflow. The outcome is a per-pixel SW ST Quality Assurance (QA) map.

\subsection{Split Window Algorithm}
\label{method:sw}
\begin{figure*}[b]
\centering
\includegraphics[width=1.0\textwidth]{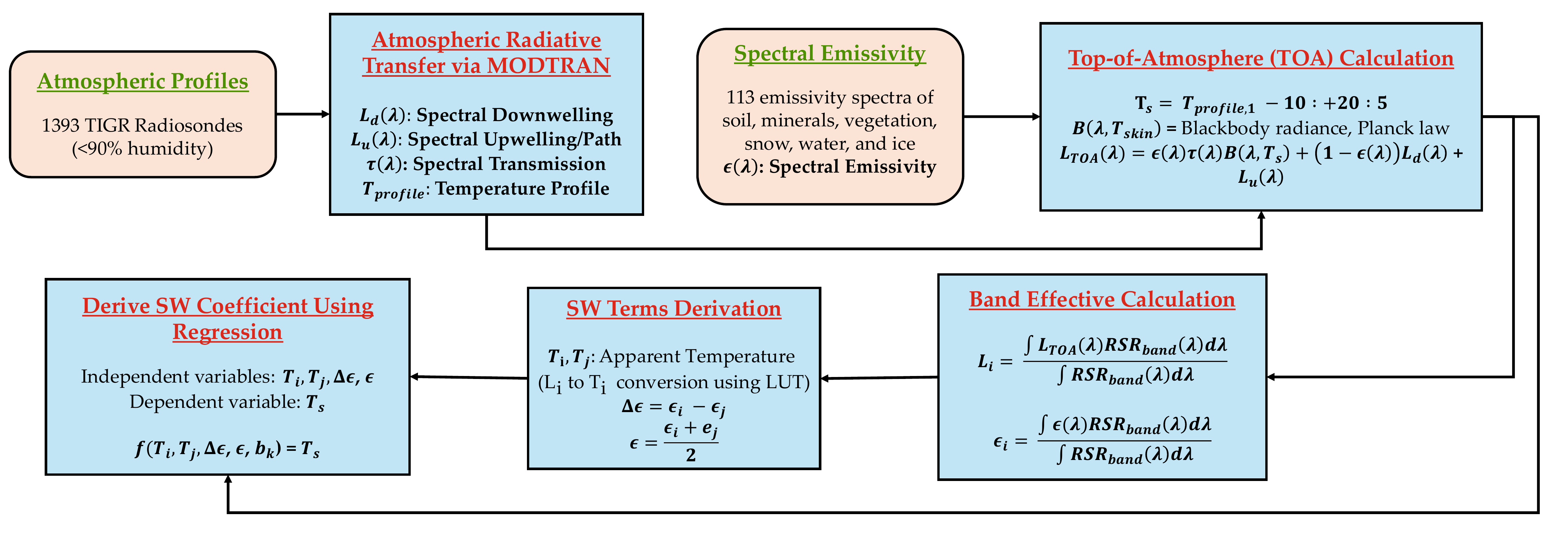}
\caption{\label{fig:methods:SWTrainingWorkflow} We leverage spectral emissivity dataset MODIS UCSB along with various atmospheric profiles from TIGR dataset forward-modeled using MODTRAN to train the SW algorithm and derive $b_k$ coefficients for L8 and L9.}
\end{figure*}

We use the version of the generalized SW algorithm that includes the quadratic term suggested by Wan (2014)~\cite{wan} introduced in Equation~\ref{eq:methods:SW-equation1} for environmental applications, i.e., we excluded spectral emissivities of man-made materials in the training process. The training spans the relevant parameters that are anticipated during data acquisition: a range of atmospheric conditions, surface temperatures, and spectral emissivities. Figure~\ref{fig:methods:SWTrainingWorkflow} presents our workflow for training the SW algorithm and we provide further explanation for each step in more detail below. 

Training the SW algorithm, i.e., deriving the b-coefficients requires band effective emissivities and apparent temperatures, as can be seen in~\ref{eq:methods:SW-equation1}. To do so, our first step in training the algorithm involves modeling the at-sensor radiance under varying atmospheric conditions. We compute the top-of-atmosphere (TOA) radiance using Equation~\ref{GoverningEq-equation2}, which combines atmospheric contributions with surface-emitted components to produce the forward-modeled radiance ($L(\lambda)$) and, ultimately, the band-effective apparent temperatures required by Equation~\ref{eq:methods:SW-equation1}

\begin{equation}
\begin{aligned}
    L(\lambda) & = \epsilon(\lambda) \tau(\lambda) B(\lambda, T_s) \\ & + (1-\epsilon(\lambda))\tau(\lambda) L_{d}(\lambda) \\ & + L_{u}(\lambda),
\end{aligned}
\label{GoverningEq-equation2}
\end{equation}

\noindent where $\epsilon(\lambda)$ is the spectral emissivity, $B(\lambda, T_s)$ is the spectral radiance emitted by a blackbody at temperature $T_s$ using Planck's equation. $\tau(\lambda)$, $L_{d}(\lambda)$ and $ L_{u} (\lambda)$ are spectral transmission, spectral downwelling radiance and upwelling radiance, respectively, all derived from MODTRAN.

We use the TIGR~\cite{tigr} atmospheric database as input to MODTRAN in order to characterize the atmospheric components of the model (see~\cite{gerace2020towards} for more information). Note that the final training environment includes 1393 TIGR profiles exhibiting a relative humidity of less than 90\%. We perform this filtering to avoid saturated atmospheric conditions. On the MODTRAN side, for each atmospheric profile, MODTRAN models spectral downwelling $L_{d}(\lambda))$, upwelling ($L_{u}(\lambda)$), and transmission ($\tau(\lambda)$). The temperature profile of the different layers of atmosphere ($T_{profile}$) are derived from TIGR profiles and input to MODTRAN. We vary the input surface temperatures to Modtran from -10~K to +20~K (in 5~K increments) about the air temperature of the temperature profile's (($T_{profile}$)) lowest layer. This is because the SW method is most suitable to use when surface and air temperatures are close~\cite{becker1990towards,wan1996generalized}.

\begin{figure}
\centering
\includegraphics[width=1.0\linewidth]{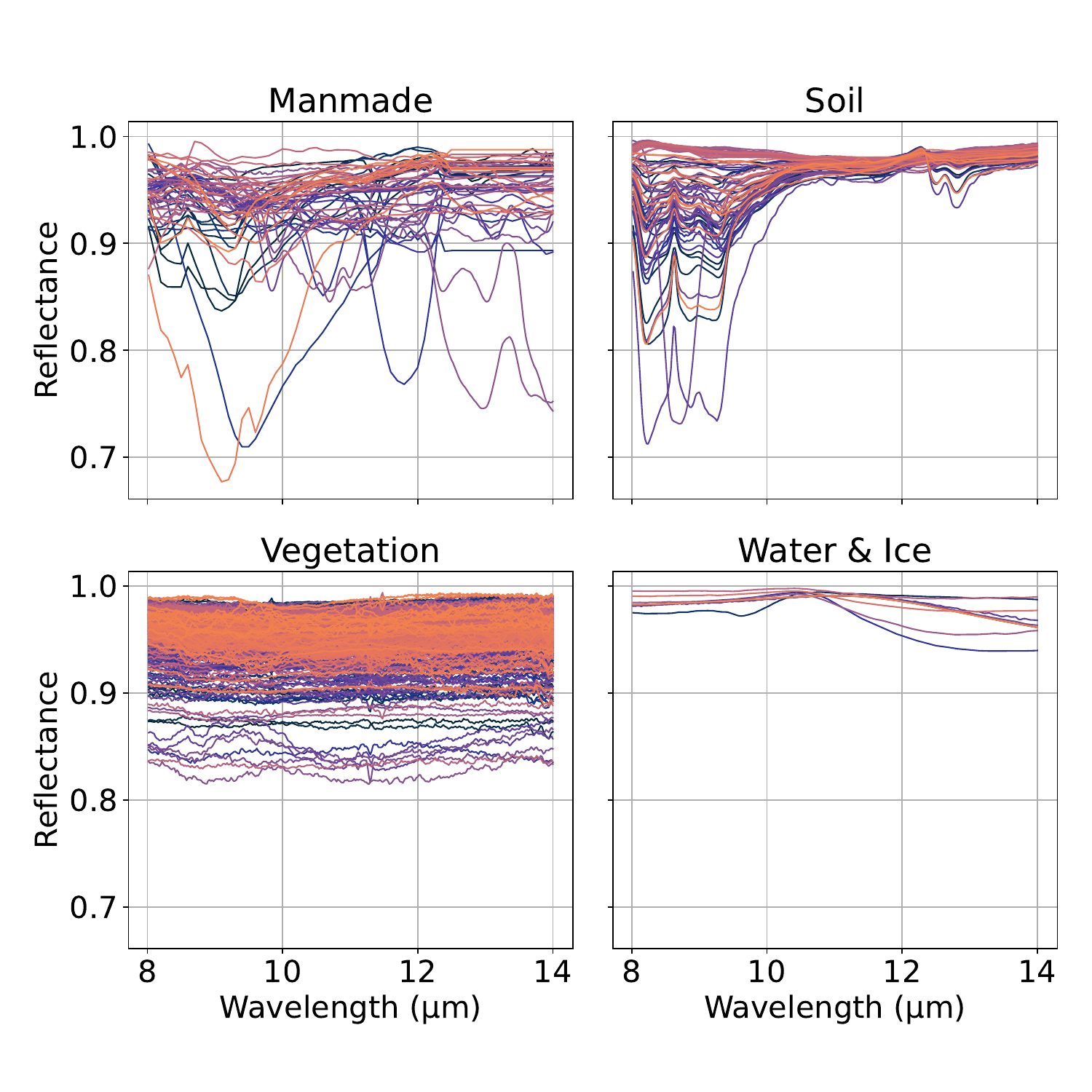}
\caption{\label{fig:methods:sw_training_emiss} MODIS UCSB emissivity database. We use 113 emissivity spectra from various materials to train the SW algorithm. }
\end{figure}

We use 113 spectral emissivities of natural materials from the MODIS UCSB emissivity database~\cite{modisemis}, i.e., vegetation, water (including snow and ice), and soils/minerals for the $\epsilon(\lambda)$ term (see Figure~\ref{fig:methods:sw_training_emiss}). 

Using the terms defined above, we first compute the top-of-atmosphere radiance ($L(\lambda)$) using Equation~\ref{GoverningEq-equation2}. Next, we derive band-effective sensor-reaching radiance and emissivity by applying the spectral response functions of Band 10 and Band 11 onboard TIRS-1 and TIRS-2. These band-effective radiances are converted to apparent temperature ($T_i$ and $T_j$) using a look-up-table. Finally, we calculate the difference in band-effective emissivities ($\Delta e$) and average band-effective emissivities ($e$) of the two bands. 

The final training data has over 1.1 million different samples (scenarios) representative of a wide range of imaging conditions for a sensor in-orbit. Finally, we develop a relationship between the independent variable terms on the right hand side of Equation~\ref{eq:methods:SW-equation1} and dependent variable ST using least-squares regression and derive the b-coefficients. Table~\ref{tab:sw_bcoeff} shows the optimized coefficients for both TIRS-1 and TIRS-2 sensors along with the corresponding RMSE values.

\begin{table}[h]
\centering
\caption{Split-Window derived b-coefficients for TIRS-1 and TIRS-2.}
\small
\begin{tabular}{cccccc}
\hline
\textbf{Satellite} & \multicolumn{4}{c}{\textbf{Coefficients}} & \textbf{RMSE [K]} \\
\hline
 \multirow{4}{*}{L9}  & \(\boldsymbol{b_{0}}\) & \(\boldsymbol{b_{1}}\) & \(\boldsymbol{b_{2}}\) & \(\boldsymbol{b_{3}}\) &  \multirow{4}{*}{0.74} \\ 
  & 2.141 & 0.994  & 0.153   & \(-0.276\) 
  & \\ 
  & \(\boldsymbol{b_{4}}\) & \(\boldsymbol{b_{5}}\) & \(\boldsymbol{b_{6}}\) & \(\boldsymbol{b_{7}}\) 
  & \\ 
  & 3.322  & 0.330  & \(-2.931\) & 0.157 
  & \\ 
\hline
 \multirow{4}{*}{L8}  & \(\boldsymbol{b_{0}}\) & \(\boldsymbol{b_{1}}\) & \(\boldsymbol{b_{2}}\) & \(\boldsymbol{b_{3}}\) &  \multirow{4}{*}{0.73} \\
  & 2.293 & 0.993  & 0.154   & \(-0.312\) 
  & \\ 
  & \(\boldsymbol{b_{4}}\) & \(\boldsymbol{b_{5}}\) & \(\boldsymbol{b_{6}}\) & \(\boldsymbol{b_{7}}\) 
  & \\ 
  & 3.719  & 0.350  & \(-3.589\) & 0.172 
  & \\ 
\hline
\end{tabular}
\label{tab:sw_bcoeff}
\end{table}

Once b-coefficients are derived, one can apply the SW Equation~\ref{eq:methods:SW-equation1}, given that the apparent temperature and the emissivity of the target material for respective TIRS bands are known. TIRS sensors do not exhibit the spectral coverage required to accurately derive band-effective emissivity~\cite{gillespie1999temperature}. As such, and to remain consistent with the SC algorithm currently being used to derive ST in Landsat's Collection-2, we instead draw on two global emissivity products—ASTER Global Emissivity Database (ASTER-GED)\cite{aster} and the Combined ASTER and MODIS Emissivity database over Land (CAMEL)\cite{camel}—to derive band-effective emissivities for TIRS Bands 10 and 11. Because ASTER-GED covers only about 96.3\% of the Earth’s land surface, gaps in its coverage would translate directly into gaps in the ST product. We therefore use CAMEL as a complementary source wherever ASTER data are unavailable, ensuring continuous emissivity input and complete ST retrieval.

For ASTER, we spectrally adjust to account for the differences in TIRS and ASTER spectral responses based on band-effective emissivities. We calculate band-effective emissivities from the 113 emissivities in the MODIS UCSB emissivity database for ASTER band 13 ($\epsilon_{aster, b13}$) and band 14 ($\epsilon_{aster, b14}$) and TIRS band 10 ($\epsilon_{b10}$) and 11 ($\epsilon_{b11}$). These paired values allow us to translate the ASTER‐GED emissivities into TIRS-2 band‐effective emissivities. Finally, we regress the TIRS data against the ASTER data according to Equations~\ref{eq: e10} and \ref{eq: e11}. Moreover, we make emissivity adjustments based on coincident reflective image data acquired from the Operational Land Imager (OLI) to update the ASTER-GED based on in-scene conditions.

\begin{align}
    \epsilon_{b10} &= c_0 \epsilon_{\text{aster}, b13} + c_1 \epsilon_{\text{aster}, b14} + c_2\label{eq: e10} \\
    \epsilon_{b11} &= c_0 \epsilon_{\text{aster}, b13} + c_1 \epsilon_{\text{aster}, b14} + c_2
    \label{eq: e11}
\end{align}

For CAMEL, we exploit its designated hinge‐point bands to derive TIRS band‐effective emissivities, but unlike ASTER, CAMEL provides only single‐wavelength emissivity values (no spectral response functions). To estimate the band‐effective emissivity for TIRS Band 10 ($\epsilon_{b10}$), we regress CAMEL’s Band 9 emissivity ($\epsilon_{camel,b9}$) and Band 11 emissivity ($\epsilon_{camel,b11}$) against known TIRS values.  Similarly, for TIRS Band 11 ($\epsilon_{b11}$), we use CAMEL’s Band 11 ($\epsilon_{camel,b11}$) and Band 12 ($\epsilon_{camel,b12}$). The resulting linear relationships are:

\begin{align}
    \epsilon_{b10} &= c_0 \epsilon_{\text{camel},b9} + c_1 \epsilon_{\text{camel},b11} + c_2\label{eq: e10_camel} \\
    \epsilon_{b11} &= c_0 \epsilon_{\text{camel},b11} + c_1 \epsilon_{\text{camel},b12} + c_2
    \label{eq: e11_camel}
\end{align}

We present the coefficient values along with the standard deviation of the fit in Table~\ref{tab:aster_coeffs} for ASTER, and in Table~\ref{tab:camel_coeffs} for CAMEL. 

\begin{table}[H]
\centering
\caption{Regression coefficients and standard deviations for the relationship between \textbf{TIRS and ASTER} band-effective emissivity using MODIS UCSB database.}
\small
\begin{tabular}{ccccc c}
\toprule
\textbf{Satellite} & & \boldmath{$c_0$} & \boldmath{$c_1$} & \boldmath{$c_2$} & \textbf{STD} \\
\midrule
\multirow{2}{*}{L9} 
& Band 10 & 0.6805 & 0.3153 & 0.0043 & 0.0006 \\
& Band 11 & -0.5825 & 1.4652 & 0.1156 & 0.005 \\
\midrule
\multirow{2}{*}{L8} 
& Band 10 & 0.5647 & 0.4254 & 0.0101 & 0.001 \\
& Band 11 & -0.5598 & 1.4464 & 0.1116 & 0.005 \\
\bottomrule
\end{tabular}
\label{tab:aster_coeffs}
\end{table}

After converting ASTER and CAMEL to TIRS band-effective emissivities, we then apply scene-specific adjustments using coincident OLI data. For ASTER, we follow the approach of Malakar et al. (2018) \cite{emisexplain}, which corrects for differences in Normalized Difference Vegetation Index (NDVI) and Normalized Difference Snow Index (NDSI) between the ASTER acquisition and the TIRS overpass.  Likewise, for CAMEL we use the adjustments described by Borbas et al. (2018) \cite{camel}, which compensate for temporal mismatches in vegetation and snow cover. These refinements ensure that our final emissivities accurately reflect the actual land-surface conditions at the moment of TIRS imaging.

\begin{table}[h]
\centering
\caption{Regression coefficients and standard deviations for the relationship between band effective \textbf{TIRS and CAMEL} database using MODIS UCSB database.}
\small
\begin{tabular}{ccccc c}
\toprule
\textbf{Satellite} & & \boldmath{$c_0$} & \boldmath{$c_1$} & \boldmath{$c_2$} & \textbf{STD} \\
\midrule
\multirow{2}{*}{L9} 
& Band 10 & 0.6521 & 0.2961 & 0.0506 & 0.0020 \\
& Band 11 & 0.1791 & 0.7712 &  & 0.0477 \\
\midrule
\multirow{2}{*}{L8} 
& Band 10 & 0.5546 & 0.3848 & 0.0592 & 0.0022 \\
& Band 11 & 0.2045 & 0.7470 & 0.04655 & 0.0025 \\
\bottomrule
\end{tabular}
\label{tab:camel_coeffs}
\end{table}

\subsection{Split Window Uncertainty Workflow}
In this section, we develop and discuss an operational workflow to generate uncertainty maps to complement the SW ST product. We designed this workflow with the intention of removing the DTC dependence that currently contaminates the Landsat Level-2 ST uncertainty maps, recalling Figure~\ref{fig:dtc_false_positives} (right).

We begin by performing standard error propagation as outlined in Equation~\ref{equation:sw_uncertainty}, to serve as a baseline for the operational uncertainty workflow. 

Looking closer at terms in Equation~\ref{equation:sw_uncertainty}, $\rho(T)_{ij} = 0.999$ and $\rho(\epsilon)_{ij} = 0.7$ are the correlation coefficients between the band-effective apparent temperatures modeled from the TIGR profiles and the 113 band-effective emissivities used in the SW algorithm development. The remaining terms are:

\begin{enumerate}\label{twp_calculation}

\item Algorithm ($\delta b$): accounts for the uncertainty in the SW algorithm that incorporates atmosphere and TPW due to the fit ($b_k$ coefficients). The baseline approach treats this uncertainty as scalar values, using the RMSE values from Table~\ref{tab:sw_bcoeff}. In contrast, our approach incorporates TPW by modeling the error due to $b_k$ coefficients as a function of TPW, i.e., $\delta b(TPW)$. This allows the uncertainty to vary dynamically with atmospheric water vapor, where higher TPW values correspond to greater uncertainty. We further discuss the uncertainty of the $b_k$ coefficients as a function of TPW in Section~\ref{tpw_error_modelling}.

\item Sensor Noise ($\delta T_i$): the sensor noise for TIRS-1 is based on a stray light analysis performed by Gerace et al. 2017~\cite{gerace2017derivation} where the residuals of reference data vs. corrected TIRS data are used to replace the reported Noise Equivalent Temperature Difference (NEdT) of the system (the residuals are bigger than the NEdT of the sensor). On the other hand, ($\delta T_i$) for TIRS-2 uses NEdT directly. We represent the sensor noise values in below: 

\begin{table}[H]
\centering
\caption{Landsat 8 and 9 sensor noise terms from the stray light analysis.}
\small
\begin{tabular}{ccc}
\toprule
\textbf{Satellite} & \(\boldsymbol{\delta T_{10}}\) (K) & \(\boldsymbol{\delta T_{11}}\) (K) \\
\midrule
L9 & 0.10 & 0.10 \\
L8 & 0.15 & 0.20 \\
\bottomrule
\end{tabular}
\end{table}

\item Emissivity ($\delta \epsilon_i$): we calculate the uncertainty in emissivity for each band, $\delta \epsilon_i$, as denoted in Equation~\ref{Eqn:emissivity_uncertainty}.

\begin{figure*}[t]
\begin{equation}
\label{Eqn:emissivity_uncertainty} 
\delta \epsilon_i = \sqrt{
    \begin{aligned} & \delta c^2 + \left(\frac{\partial\epsilon_{i}}{\partial\epsilon_{\text{aster},13}}\delta \epsilon_{\text{aster},13} \right)^2 
    + \left(\frac{\partial\epsilon_{i}}{\partial\epsilon_{\text{aster},14}}\delta \epsilon_{\text{aster},14} \right)^2 \\&
    + 2\rho(\epsilon_{\text{aster},13},\epsilon_{\text{aster},14}) \frac{\partial\epsilon_{i}}{\partial\epsilon_{\text{aster},13}}\frac{\partial\epsilon_{i}}{\partial\epsilon_{\text{aster},14}} \delta \epsilon_{\text{aster},13} \delta \epsilon_{\text{aster},14}
    \end{aligned}
    }
\end{equation} 
\end{figure*}

For brevity we only explain the ASTER‐based uncertainty propagation; the CAMEL formulation is identical but uses its own hinge‐point bands. Referring to Equation~\ref{Eqn:emissivity_uncertainty}, $\delta c$ denotes the standard deviation of the target band fit (Tables \ref{tab:aster_coeffs} and \ref{tab:camel_coeffs}), while
$\displaystyle \frac{\partial \epsilon_i}{\partial \epsilon_{\text{aster},13/14}}$
is the partial derivative of the target band emissivity $\epsilon_i$ (from Eqs.~\ref{eq: e10} and \ref{eq: e11}) with respect to ASTER Band 13 or 14 emissivities. The per‐pixel uncertainties $\delta\epsilon_{\text{aster},13/14}$ come directly from the ASTER‐GED standard‐deviation maps, and the correlation between the two ASTER band‐effective emissivities is
$\rho\bigl(\epsilon_{\text{aster},13},\,\epsilon_{\text{aster},14}\bigr) = 0.8923$.
For CAMEL, the corresponding correlations are
$\rho\bigl(\epsilon_{\text{camel},9},\,\epsilon_{\text{camel},11}\bigr) = 0.8774$,
\quad
$\rho\bigl(\epsilon_{\text{camel},11},\,\epsilon_{\text{camel},12}\bigr) = 0.7337$.
The same framework described above applies to the CAMEL‐derived bands.

\end{enumerate}

\label{tpw_error_modelling}
\subsubsection{Capturing SW Uncertainty Through Atmospheric Water Vapor via MODTRAN
}

The objective of this section is to account for error in the SW algorithm by establishing a relationship between algorithmic error and atmospheric water vapor. To build a representative dataset across different scenarios, we follow the same workflow introduced in Section~\ref{method:sw} for training the SW algorithm, with one key modification: we vary atmospheric water vapor by adjusting the MODTRAN water vapor scaler which allows us to simulate different levels of water vapor within a given atmospheric profile. As a result, the variability in algorithm error ($\text{SW\_ST} - \text{LST}$) for each profile is primarily driven by changes in emissivity spectra, skin temperature ($T_s$), and TPW.

\begin{figure}
    \centering
    \includegraphics[width=0.8\linewidth]{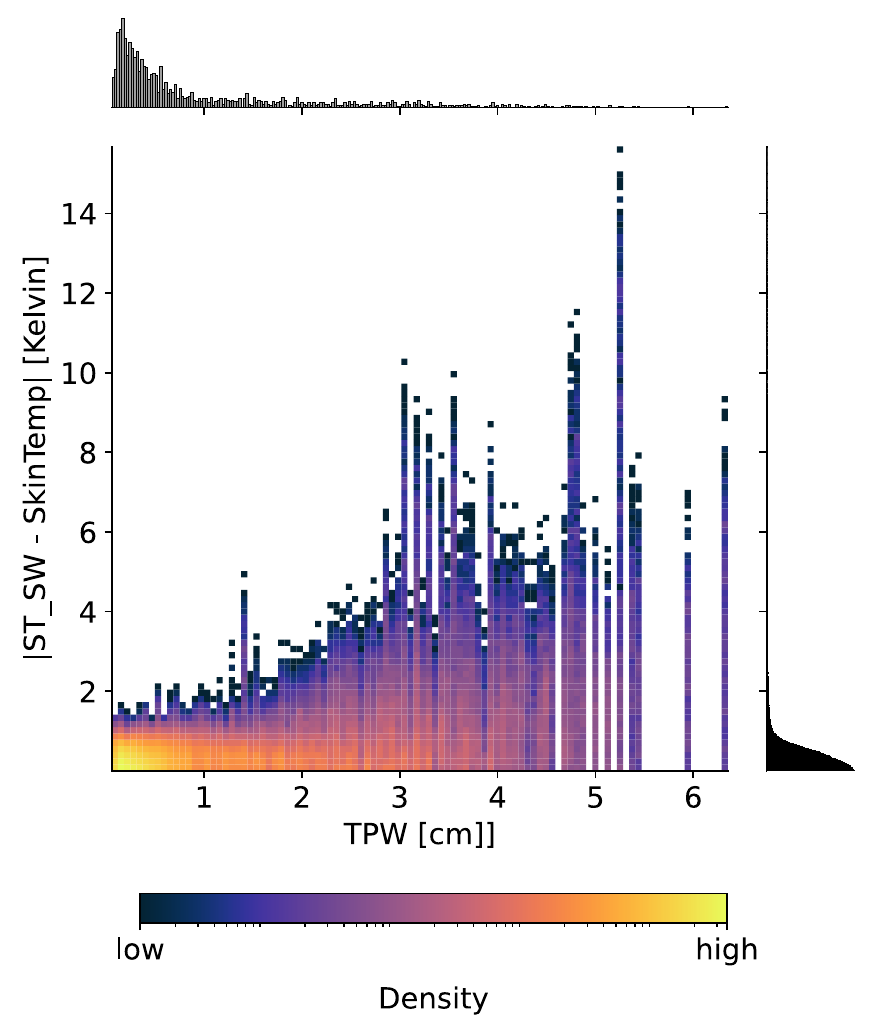}
    \caption{The scatter plot of algorithmic error vs. TPW. We can see that the higher TPW corresponds to higher algorithmic error.}
    \label{fig:swerror_tpw}
\end{figure}

Figure~\ref{fig:swerror_tpw} shows the scatter plot of algorithmic error vs. TPW. The data are most concentrated in regions of low TPW and low error, while TPW values above 6 cm become increasingly sparse, with only two TIGR atmospheric profiles present. One could argue that this imbalance might bias the relationship between error and TPW. To evaluate this, we perform undersampling in the dense low-TPW region ($\text{TPW} \in [0, 2]$ cm). However, the resulting fit from the undersampled dataset showed minimal differences from the original dataset. Therefore, for simplicity, we proceed with the fit derived from the full dataset.

\begin{figure}
    \centering
    \includegraphics[width=1\linewidth]{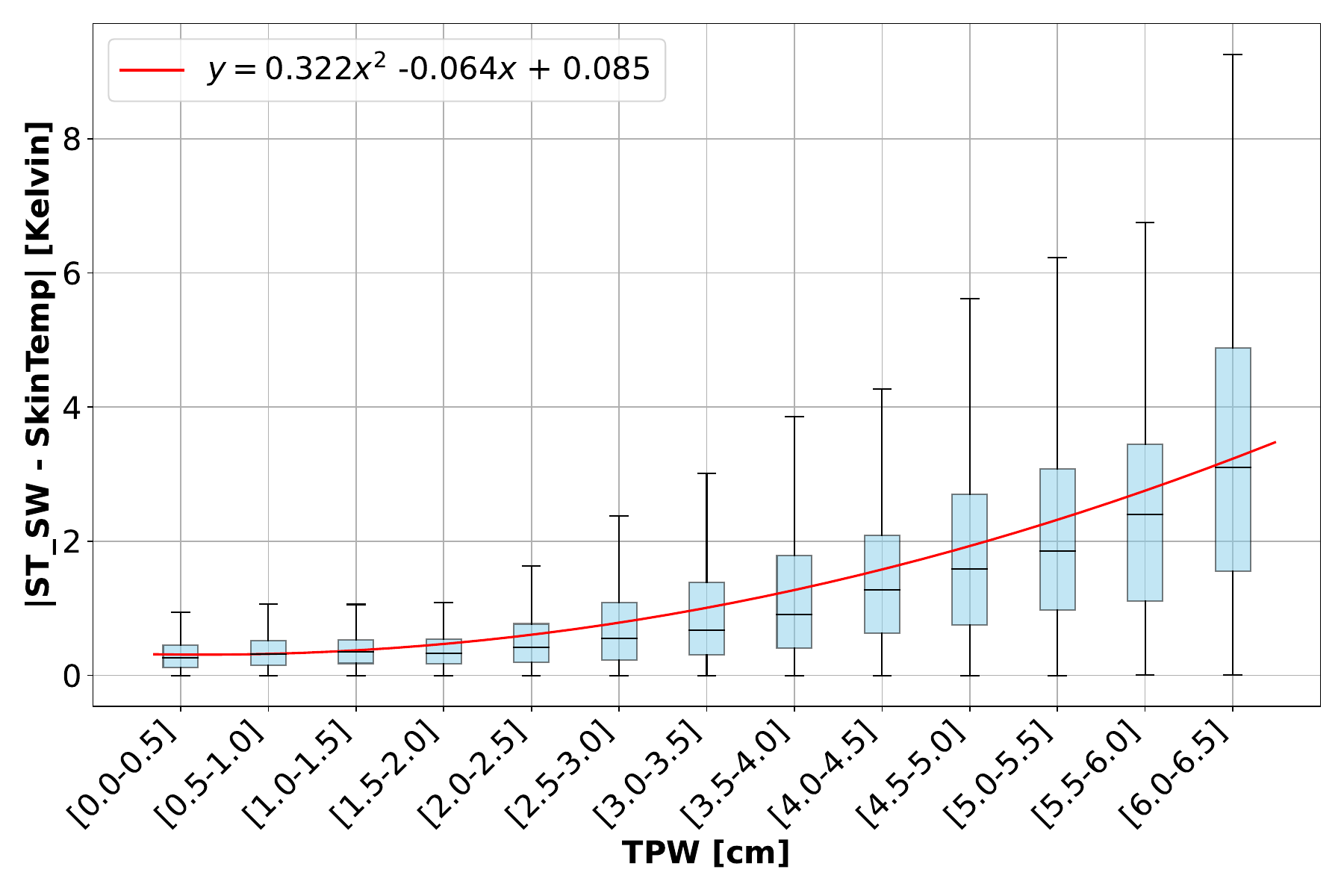}
    \caption{SW algorithmic error vs. TPW in terms of boxplots. Note that with higher TPW values we observe larger variability in terms of algorithmic error. We establish a quadratic fit between error and TPW.}
    \label{fig:swerror_tpw_relationship}
\end{figure}

Figure~\ref{fig:swerror_tpw_relationship} presents a boxplot of algorithmic error across TPW bins. We use data from Figure~\ref{fig:swerror_tpw} to derive the quadratic fit displayed in Figure~\ref{fig:swerror_tpw_relationship}. Albeit loosely correlated, this relationship enables the approximation of algorithmic error ($\delta b$) from a user-specified TPW value. The established relationship better captures the algorithmic uncertainty based on variable TPW, compared to a fixed scalar RMSE of b-coefficients fit.  

In the next subsection, we present an operational machine learning approach for estimating TPW using L9 TIRS bands.

\subsection{Operational TPW Estimation Using L9 TIRS}

\begin{figure}[b]
\centering
\includegraphics[width=1\linewidth]{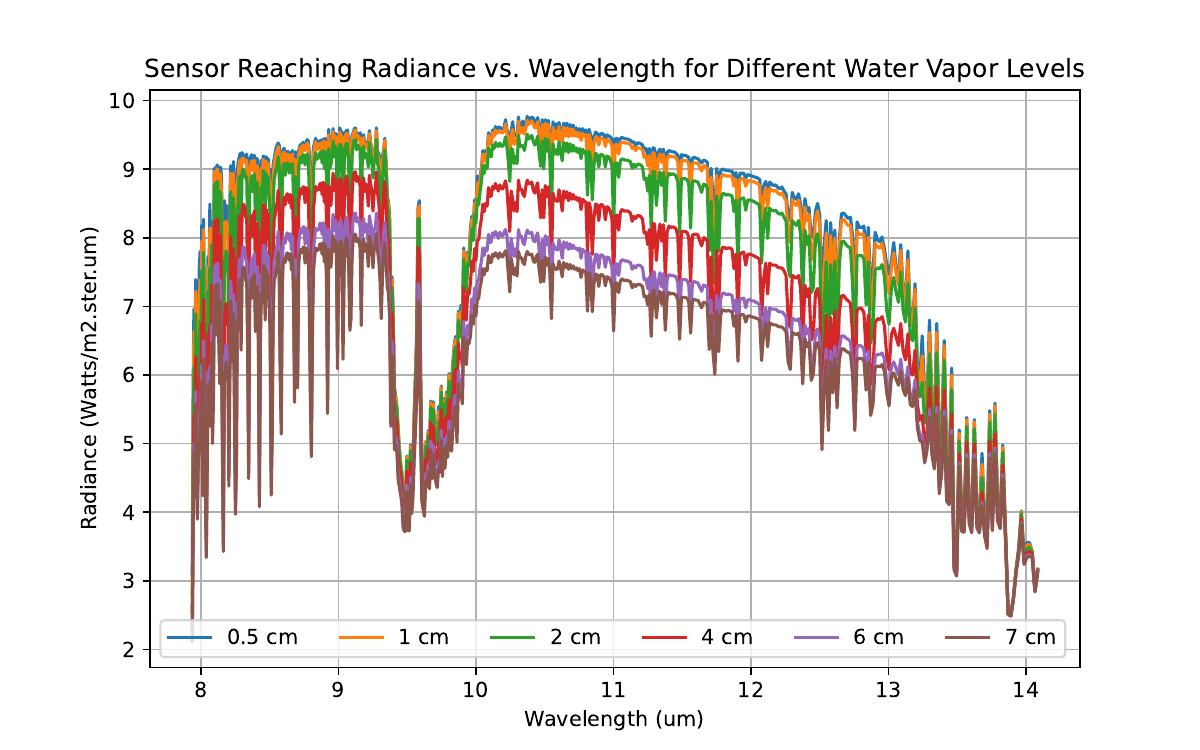}
\caption{\label{fig:methods:ThermalBehaviorVsTPW} At-sensor radiance for varying levels of Total Precipitable Water (TPW). Note the drop in the overall magnitude of the signal with the increase in TPW levels.}
\end{figure}

As mentioned, in order to retrieve the algorithmic error passed to the uncertainty workflow, one would need an estimate of per-pixel TPW. In order to understand the impact of water vapor in the atmosphere on sensor-reaching radiance, we investigate the relationship between TPW and sensor-reaching radiance using MODTRAN simulations at varying water vapor levels (see. Figure~\ref{fig:methods:ThermalBehaviorVsTPW}). The figure shows that there exists an inverse relationship between atmospheric water vapor and the TOA radiance in the thermal infrared region.

An estimate of atmospheric water vapor can increase the accuracy of split window uncertainty estimation. TIRS-1 and TIRS-2 do not have the proper band to accurately estimate atmospheric water vapor, but MODIS Terra is equipped to do so. In fact, MODIS05\_L2 is a TPW product derived by two-channel or three-channel ratio approaches~\cite{gao2003water}. Terra-MODIS TPW product has been validated against Aerosol Robotic Network (AERONET) perceptible water with a global average uncertainty of MAE = $0.402$ cm, establishing the product as a suitable candidate in terms of accuracy for the mentioned modeling task~\cite{bright2018climatic}. Moreover, the L9 and MODIS-Terra satellites have near-coincidental overpasses, with MODIS approximately 10 minutes behind L9. This temporal proximity allows us to use MODIS-derived TPW as reference for developing a predictive model using L9 TIRS bands. We define the mapping function between TIRS 2 B10 and B11 with the MODIS-TPW product as:

\begin{equation}
f(\text{L}_{B10}, \text{L}_{B11}) = \widehat{\text{TPW}}
\label{eq:tpw_mapping}
\end{equation}

\noindent where $\text{L}_{Bi}$ represents the TOA radiance for L9 B10 and B11. 

The astute reader might question why MODIS is used instead of a radiative transfer model (RTM) like MODTRAN. While MODTRAN can simulate the atmosphere, using MODIS offers key advantages: it better reflects real-world conditions, accounts for heterogeneous atmospheres, captures sensor noise and mixed pixels, and provides a large dataset that supports balanced training and strong model performance. In the following sections, we delve into the datasets and models used to derive the mentioned mapping function.

\subsubsection{Dataset}
It is important to note that on October 12, 2022, the MODIS Terra satellite underwent a Constellation Exit Maneuver, which removed it from its original orbit and nominal constellation alignment~\cite{modis_exit}. As a result, we restrict our data collection to coincident L9 TIRS and MODIS05\_L2 observations from early 2022 through the beginning of October 2022.

\begin{figure}[h]
    \centering
    \includegraphics[width=\linewidth]{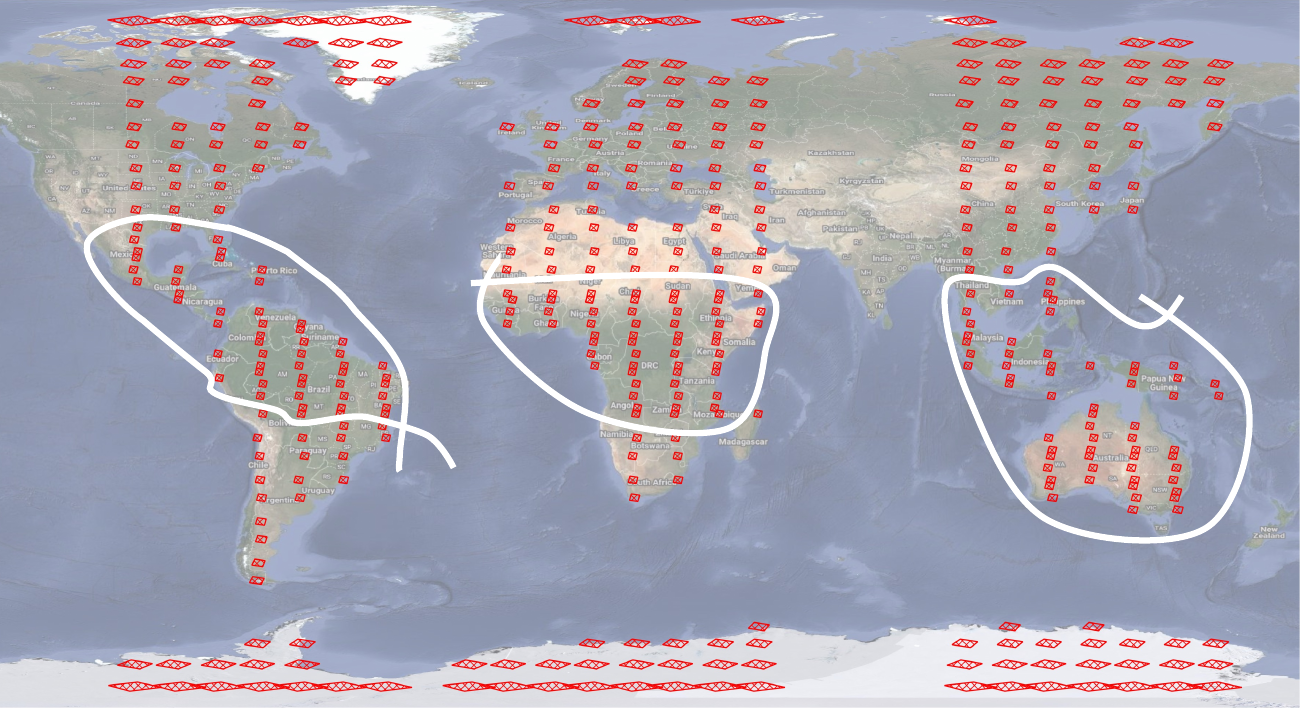}
    
    \vspace{1em}
    
    \includegraphics[width=\linewidth]{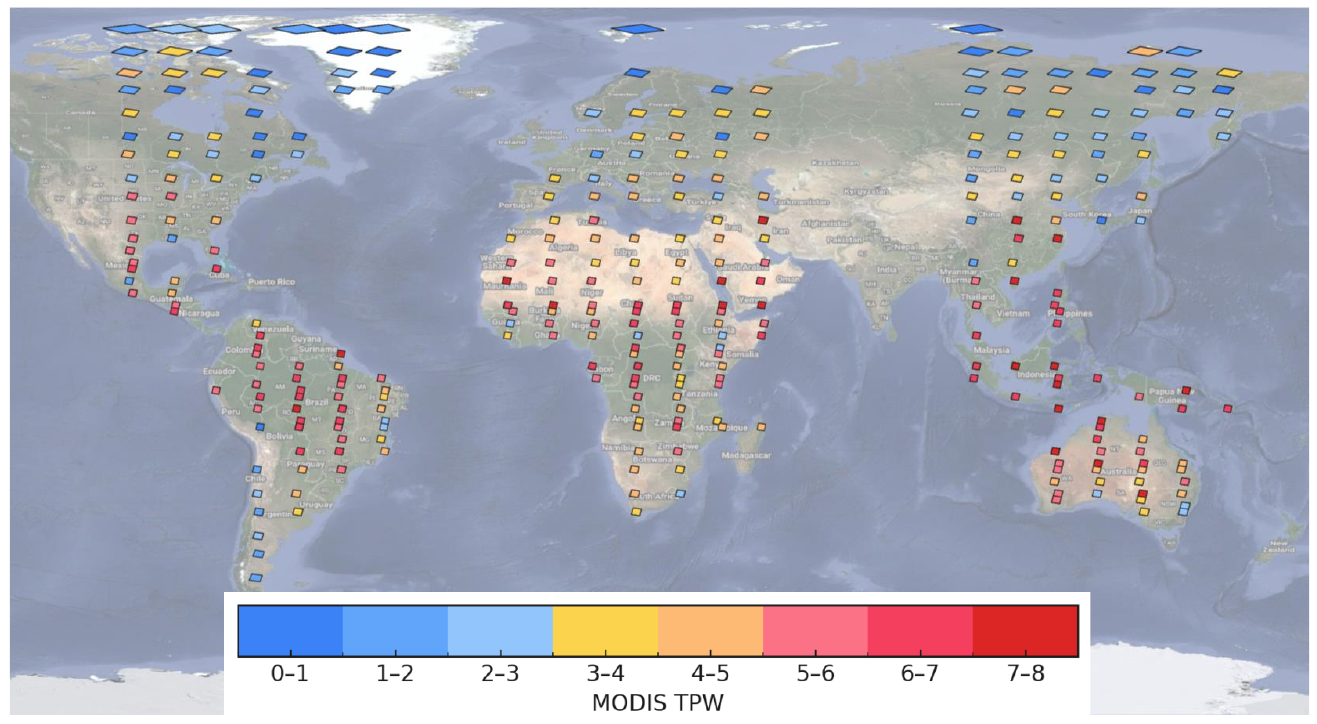}
    
    \caption{\textit{Top}: Scene Path/rows highlighted in red were used to collect the original dataset of 55 million samples. Polygons indicate regions with higher TPW values where a doubled sampling rate is applied. \textit{Bottom}: Corresponding Path/rows for the balanced, randomly undersampled dataset consisting of 7,990 samples. The colorbar represents MODIS TPW values, binned across the 0–8 range.}
    \label{fig:wrs2sample}
\end{figure}

Our data collection strategy is based on the World Reference System-2 (WRS-2) grid and the available Path/Row combinations with the aim of capturing a broad-range of atmospheric conditions. Because near-equatorial scenes typically contain more clouds, we apply denser sampling to build a more balanced dataset. In humid equatorial regions, we sample at 10-degree intervals longitudinally and 2.5-degree intervals latitudinally, see Figure~\ref{fig:wrs2sample} (top). Outside these regions, we increase the latitudinal sampling interval to 5 degrees. Due to the high frequency of cloud cover in these areas and the need for extensive filtering and masking to ensure cloud-free pixels, this dense sampling approach is necessary to obtain sufficient data in high-TPW regions.

\begin{figure*}
    \centering
    \includegraphics[width=0.95\linewidth]{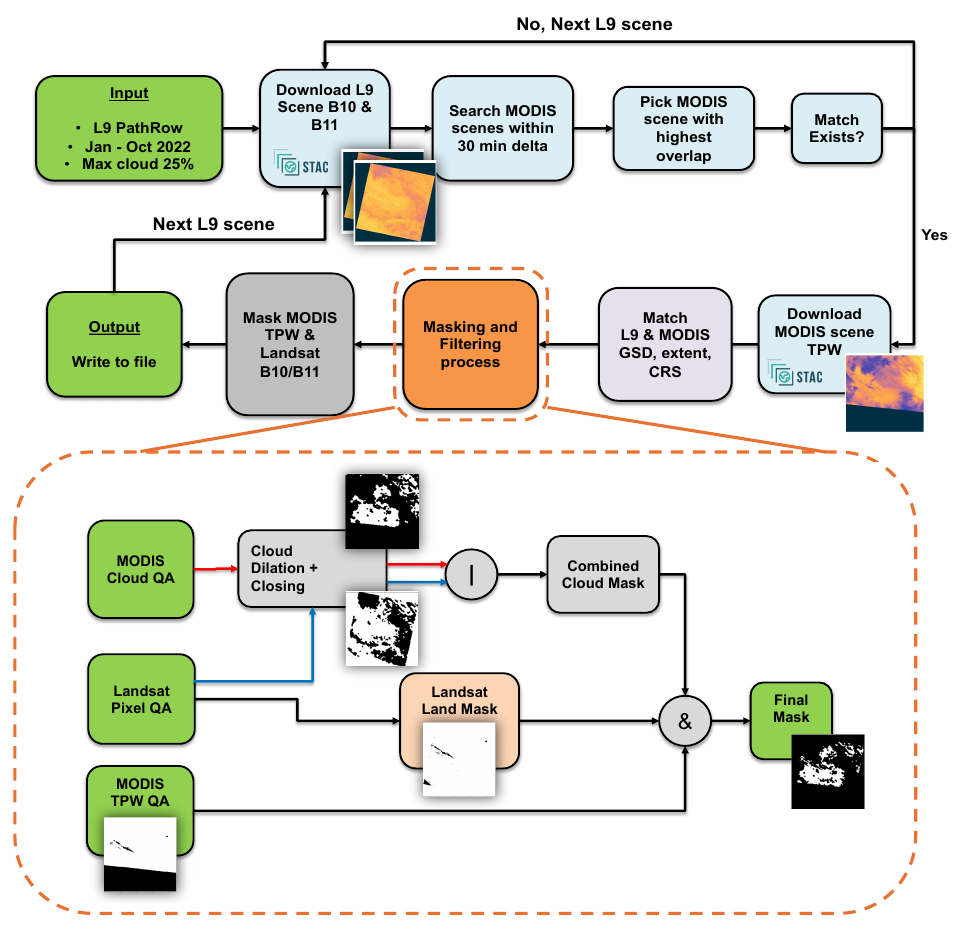}
    \caption{Data collection workflow for acquiring coincidental Landsat B10/B11-MODIS TPW data.}
    \label{fig:tpw_data_collection_workflow}
\end{figure*}

We present the data collection workflow in Figure~\ref{fig:tpw_data_collection_workflow}. This workflow describes the use of the PySTAC Application Programming Interface (API)~\cite{stac_spec} to download coincidental L9 and MODIS TPW scenes for the period between January and October 2022, with a maximum allowable cloud coverage of 25\%. We select MODIS scenes acquired within 30 minutes of each L9 overpass with the highest spatial overlap. We then spatially downsample the L9 data to match the spatial resolution of MODIS (1 km), and ensure that the spatial extent and coordinate reference systems of both datasets are consistent. We apply a thorough masking and filtering process to ensure that the collected dataset meets quality standards for modeling purposes. We use the MODIS Cloud QA, L9 QA\_PIXEL, and MODIS TPW QA bands to generate cloud masks from both the L9 QA\_PIXEL and MODIS products, then refine these intermediate mask using dilation and morphological closing to avoid including pixels near cloud edges. We merge the two MODIS and Landsat cloud masks into a single unified mask and intersect it with the L9 land mask to remove water-covered areas. We retain only high-confidence TPW pixels using the MODIS TPW QA band. Finally, we apply the combined mask to extract valid pixel trios that include L9 TOA B10 and B11 along with their corresponding TPW values from MODIS.

\begin{figure}[h]
  \centering
  \includegraphics[width=1\linewidth]{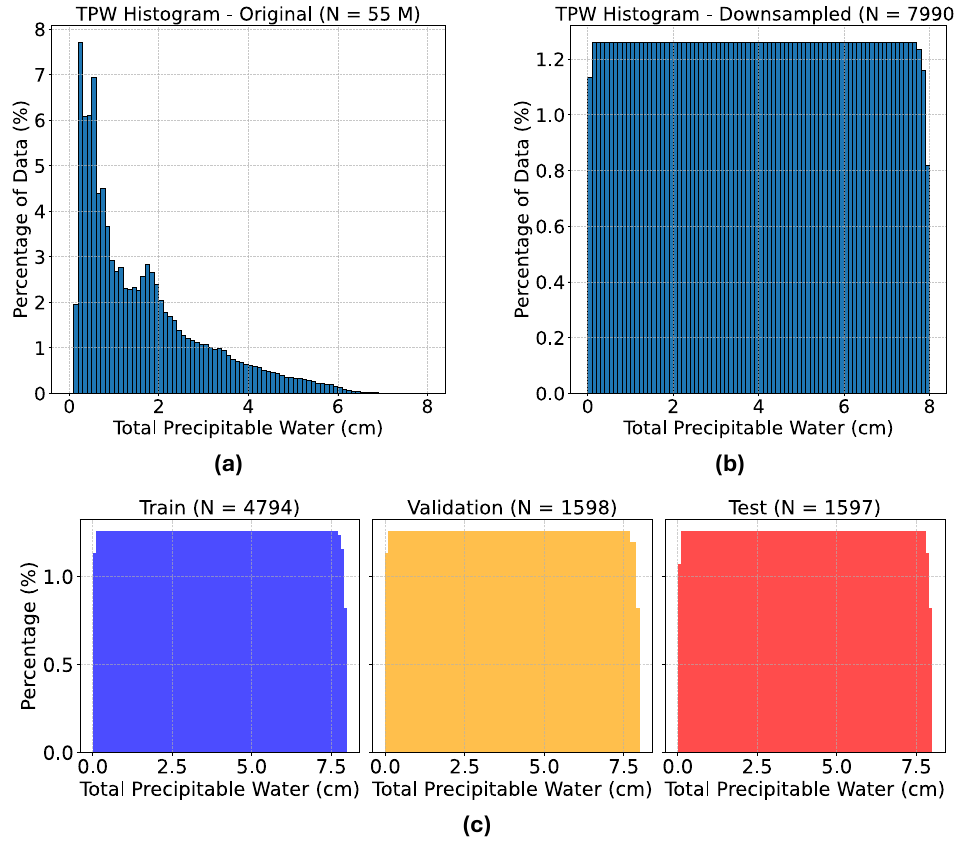}
  \caption{We present the sampling work in this figure.(a) is the original dataset TPW distribution with 55 Million samples, we perform random undersampling on the original dataset and generate (b) with 7990 samples, we partition this dataset into three balanced dataset (c) training, validation, and testing sets .}
  \label{fig:tpw_sampling_worfklow}
\end{figure}

Our proposed workflow produces a dataset with approximately 55 million samples; see Figure~\ref{fig:tpw_sampling_worfklow} (a). However, because lower TPW values are highly overrepresented, we apply random undersampling based on the TPW distribution; Figure~\ref{fig:tpw_sampling_worfklow} (b). The undersampled dataset contains 7990 $([L_{B10}, L_{B11}], TPW)$ pairs; see Figure~\ref{fig:wrs2sample} (bottom) for a spatial visualization of the corresponding path/rows building the dataset. To train the model, tune model hyperparameters, and evaluate generalization performance, we divide the undersampled dataset into three sampled balanced partitions: training, validation, and testing, using a 3:1:1 ratio. The partition distributions are approximately uniform, enabling unbiased model training and eliminating the need for balanced evaluation metrics across the dataset splits; see Figure~\ref{fig:tpw_sampling_worfklow} (c).

\begin{figure}
    \centering
    \includegraphics[width=1\linewidth]{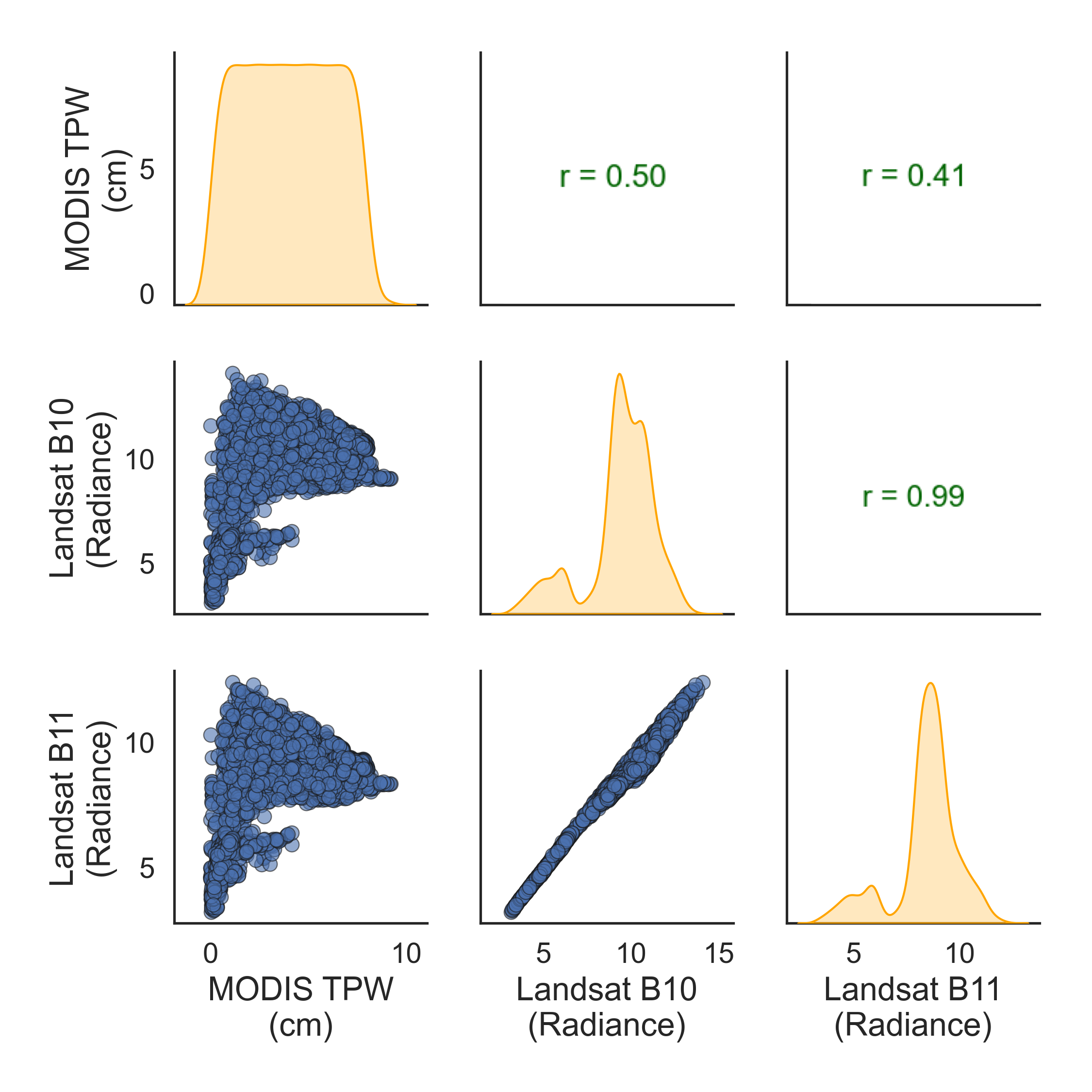}
    \caption{Pairwise plot of the independent variables, TOA B10 and B11, along with the dependent variable, TPW. The plot shows that B10 and B11 are highly correlated with each other, indicating a strong linear relationship between the two bands.}
    \label{fig:tpw_train_pairwiseplot}
\end{figure}

We present a pairwise plot of the two independent variables—$L_{B10}$ and $L_{B11}$—and the dependent variable, MODIS-derived TPW for the training partition (see Figure~\ref{fig:tpw_train_pairwiseplot}). The plot indicates that $L_{B10}$ and $L_{B11}$ are highly correlated and both show a moderate positive correlation with TPW. However, applying a regression model using only two features would likely result in high bias, underfitting, due to the limited expressiveness of the feature space. Moreover, the strong linear dependence between B10 and B11 raises concerns about multicollinearity, which can compromise the stability and interpretability of any linear regression coefficients. In the following sections, we start with our approach to modeling TPW using boosting algorithms, followed by Feature Engineering (FE) and Feature Selection (FS) step to enrich the input space.

\subsubsection{Modelling TPW via Extreme Gradient Boosting} 
Among ensemble learning models, boosting approaches are particularly effective for improved generalization on unseen data by sequentially combining multiple learners. In boosting, each successive model is trained to focus on the errors made by previous models, allowing the ensemble to effectively handle hard-to-predict instances. This iterative strategy enables the ensemble to progressively reduce both bias and variance. Extreme Gradient Boosting (XGBoost) is a widely adopted boosting algorithm that builds an ensemble of decision trees, where each tree is added in an additive manner to minimize the overall prediction error~\cite{chen2016xgboost}. It leverages gradient-based optimization and incorporates regularization of L1 (Lasso) and L2 (Ridge) to mitigate overfitting, an inherent challenge in decision tree models. Additionally, XGBoost applies tree pruning to control model complexity, which further enhances generalization and reduces the risk of overfitting. XGBoost is lightweight, GPU-accelerated, and highly efficient, boasts seamless compatibility across various programming languages and hardware platforms, facilitating smooth integration with Earth Resources Observation and Science (EROS) workflow. In this study, we use XGBoost for the purpose of modeling the dataset introduced in the previous section.

We define the feature space $X \in \mathbb{R}^{n \times m}$ as a matrix containing TOA radiance-level feature information from L9 TIRS-2 bands, where $n$ is the number of samples and $m$ is the number of features used. The corresponding target variable $Y \in \mathbb{R}^{n \times 1}$ is the MODIS-derived TPW values. These matrices can be represented as:

\begin{equation*}
    X =
\begin{bmatrix}
x_{1,1} & x_{1,2} & \cdots & x_{1,m} \\
x_{2,1} & x_{2,2} & \cdots & x_{2,m} \\
\vdots  & \vdots  & \ddots & \vdots  \\
x_{n,1} & x_{n,2} & \cdots & x_{n,m}
\end{bmatrix},
\quad
Y =
\begin{bmatrix}
y_1 \\
y_2 \\
\vdots \\
y_n
\end{bmatrix}
\end{equation*}

\noindent The goal is to learn the predictive mapping function :

\begin{equation}
F(X, \theta) \approx \hat{Y}    
\end{equation}

where $\hat{Y} \in \mathbb{R}^{n \times 1}$ is the model’s estimated TPW and $\theta$ denotes the parameters learned during training. To determine optimal parameters, XGBoost builds trees sequentially, allowing gradients to guide the optimization of the loss function. The prediction at iteration/tree $t$ as $\hat{y}_i^{(t)}$ given features space $x$ and sample $i$ is dependent on prediction at previous iteration/tree $t-1$ as $\hat{y}_i^{(t)}$ and the newly added tree:

\begin{equation}
\hat{y}_i^{(t)} = \sum_{k=1}^{t} f_k(x_i) = \hat{y}_i^{(t-1)} + f_t(x_i)
\label{eq:additive_model}
\end{equation}

\noindent XGBoost minimizes the following regularized objective function (for regression task) as:

\begin{equation}
\text{obj}^{(t)} = \sum_{i=1}^{n} \left( y_i - \left( \hat{y}_i^{(t-1)} + f_t(x_i) \right) \right)^2 + \sum_{k=1}^{t} \omega(f_k)
\label{eq:objective}
\end{equation}

\noindent where the first term represents the squared residuals, and the second term $\omega(f_k)$ penalizes model complexity (see~\cite{chen2016xgboost} for detail). 

We perform training on XGBoost using the training data, tune the hyperparameters using the validation set, and evaluate the performance of the model on the test set. We benchmark the performance of XGBoost against several other approaches, including linear regression, Partial Least Squares Regression (PLSR;~\cite{geladi1986partial}), and Multi-Layer Perceptron (MLP). In the following section, we describe our approach to feature engineering and feature selection — an important step due to the small number of input features increasing the risk of underfitting (high bias).

\subsubsection{Feature Engineering and Feature Selection} Due to the limited number of input features, there is a high likelihood of underfitting, high bias, in the regression task. To mitigate this, we perform feature engineering on the two available TOA-level bands, $L_{10}$ and $L_{11}$. This step includes a series of nonlinear transformations applied to each band individually, as well as in combination (see Table~\ref{table:feature_eng}). These transformations include division, normalized difference indexing, logarithmic, power, and sinusoidal functions. This process expands the feature space to 23 total number of features, providing a richer representation of the features for the modeling task.

\begin{table*}[h]
\centering
\caption{Summary of engineered features derived from bands $L_i$ and $L_j$ ($i,j \in \{10,11\}$), designed to enhance the predictive power of the input data. For brevity, expressions such as $\log(L_i / L_j)$ implicitly include their counterparts (e.g., $\log(L_j / L_i)$), even if not explicitly listed.}
\small
\resizebox{\textwidth}{!}{
\begin{tabular}{lll}
\toprule
\textbf{Feature Type} & \textbf{Transformation} & \textbf{Description} \\
\midrule
Raw Bands & $L_i$, $L_j$ & Original band radiance values \\
\hline
Ratios and Products & $L_i / L_j$, $L_i \cdot L_j$ & Captures relative differences \\
\hline
Logarithmic & $\log(L_i)$, $\log(L_i / L_j)$, $\log(L_i \cdot L_j)$ & Emphasizes proportional changes \\
\hline
\multirow{3}{*}{Powers} & $L_i^2$, $L_i^3$ & Amplifies variation\\
       & $(L_i / L_j)^2$, $(L_i / L_j)^3$ & Nonlinearity emphasizing ratio\\
       & $(L_i \cdot L_j)^2$, $(L_i \cdot L_j)^3$ & Nonlinearity\\
\hline
Roots & $(L_i / L_j)^{0.5}$, $(L_i \cdot L_j)^{0.5}$ & Shrinks large values\\
\hline
Sinusoidal & $\sin(L_i)$, $\cos(L_j)$ & Nonlinear periodic patterns\\
\hline
Normalized Difference Index & $\frac{L_i - L_j}{L_i + L_j}$ & Highlights contrast and normalizes dynamic range \\
\bottomrule
\end{tabular}
}
\label{table:feature_eng}
\end{table*}

On the other hand, an increase in the number of features elevates the risk of overfitting, high variance. This highlights the classic bias vs. variance trade-off, where the number of features directly influences the error in the modeling task~\cite{ziegel2003elements}. To address this, we employ a meta-heuristic optimization algorithm—Ant Colony Optimization (ACO)—adapted for feature selection. ACO offers a simple and robust framework and is computationally less expensive compared to other meta-heuristic algorithms, such as Particle Swarm Optimization~\cite{selvarajan2019comparative}. This wrapper-based approach evaluates model performance on the validation set to identify the optimal subset of features. We implement ACO using the \texttt{Jostar} library in Python and conduct feature selection across subsets containing 2 to 10 features~\cite{hassanzadeh2021broadacre}; \url{https://github.com/amirhszd/jostar}. 

\subsubsection{Training Configuration \& Evaluation Metrics}. \\
We run XGBoost with 100 estimators for the hyperparameter tuning step on the validation set and 1000 estimators at the training stage, ensuring sufficient training iterations until convergence. We train all other benchmark models with default argument values.

For the purpose of evaluating the performance across models and data partitions, we employ three standard regression metrics: Mean Absolute Error (MAE), coefficient of determination ($R^2$), and standard deviation (STD) of residuals. We use these metrics in their standard form, as datasets are balanced and do not require weighting or correction for data imbalance.




In the following results section, we present results of TPW modeling along with validation efforts using real satellite imagery.

\section{Results and Discussion}

In this section, we first present the results of TPW modeling, followed by validation outcomes based on the updated uncertainty workflow described earlier. 

\subsection{TPW Modeling}

We present the results of the TPW modeling performance from various models: LR, PLSR, MLP, linear XGBoost, and XGBoost for both training and testing sets, as shown in Table~\ref{tab:tpw_model_comparison}. Here we evaluate performance from three perspectives, overall predictive ability and capacity to capture nonlinear relationships, generalization behavior, and the impact of FE and FS.

\begin{table*}[b]
\centering
\caption{Comparison of TPW modeling performance across different models and feature sets.}
\small
\begin{tabular}{ll|ccc|ccc}
\toprule
\textbf{Model} & \textbf{Features} & \multicolumn{3}{c|}{\textbf{Training}} & \multicolumn{3}{c}{\textbf{Testing}} \\
 & & $R^2 \uparrow$ & MAE $\downarrow$ & STD $\downarrow$ & $R^2 \uparrow$ & MAE $\downarrow$ & STD $\downarrow$ \\
\midrule
LR & 2: L10, L11 & 0.71 & 1.01 & 1.25 & 0.69 & 1.04 & 1.28 \\
LR + \textcolor{blue}{FE} & 23 & 0.88 & 0.58 & 0.82 & 0.87 & 0.59 & 0.82 \\
\hline
PLSR & 2: L10, L11 & 0.71 & 1.01 & 1.25 & 0.69 & 1.04 & 1.28 \\
PLSR + \textcolor{blue}{FE} & 23 & 0.87 & 0.59 & 0.82 & 0.87 & 0.59 & 0.83 \\
\hline
MLP & 2: L10, L11 & 0.88 & 0.58 & 0.81 & 0.88 & 0.59 & 0.81 \\
MLP + \textcolor{blue}{FE} & 23 & 0.89 & 0.55 & 0.78 & 0.88 & 0.56 & 0.79 \\
\hline
Linear XGBoost & 2: L10, L11 & 0.62 & 1.21 & 1.42 & 0.61 & 1.24 & 1.44 \\
Linear XGBoost + \textcolor{blue}{FE} & 23 & 0.86 & 0.63 & 0.86 & 0.86 & 0.63 & 0.86 \\
\hline
XGBoost & 2: L10, L11 & 0.80 & 0.81 & 1.04 & 0.71 & 0.97 & 1.25 \\
XGBoost + \textcolor{blue}{FE} & 23 & 0.96 & 0.31 & 0.44 & 0.87 & 0.56 & 0.82 \\
\textbf{XGBoost + \textcolor{blue}{FE + FS}} & \textbf{3} & \textbf{0.93} & \textbf{0.44} & \textbf{0.60} & \textbf{0.89} & \textbf{0.54} & \textbf{0.78} \\
\bottomrule
\end{tabular}
\label{tab:tpw_model_comparison}
\end{table*}

\begin{figure}
    \centering
    \includegraphics[width=1\linewidth]{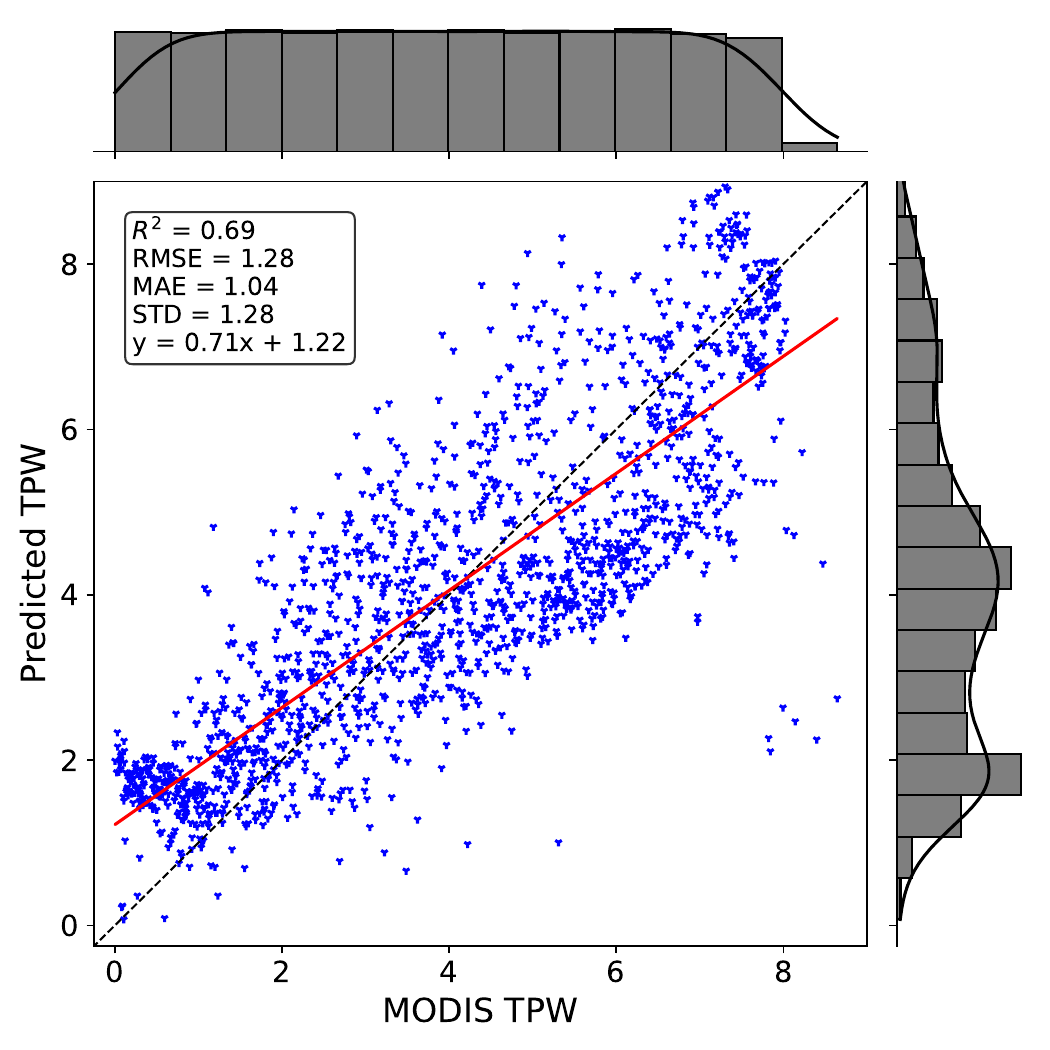}
    \caption{\textbf{LR with 2 features (L10, L11):} The regression plot shows the model's inability to capture nonlinear patterns in the data.}
    \label{fig:tpw_lr_regression}
\end{figure}

LR and PLSR exhibit lower performance compared to other models in capturing the complexity of the data, with testing MAE values around 1 cm. This outcome is expected, given their assumption of linear relationships between variables. As shown in Figure~\ref{fig:tpw_lr_regression}, both LR using only two features ($L_{10}$ and $L_{11}$) and linear XGBoost fail to capture nonlinear nature of data, demonstrating that models constrained by linearity are limited in characterizing the underlying patterns of this dataset.

Among nonlinear models, MLP and XGBoost show improved performance over linear models when using only two features ($L_{10}$ and $L_{11}$). However, MLP requires longer training time due to its lack of GPU acceleration, which also extends the time needed for hyperparameter tuning. Despite this, MLP outperforms XGBoost with two features, achieving a lower testing MAE (0.59 vs. 0.97 cm). XGBoost, which relies on iterative feature thresholding through tree-based splits, is less effective when feature space is sparse. With only two input features, the model has limited opportunities to explore complex decision boundaries, reducing its ability to fully capture nonlinearity.

Next, we enhance these models by introducing FE to enrich the input space, which improves the performance across all models. However, for XGBoost + FE, we observe a large gap between training and testing MAE — training MAE is nearly half of those for testing — highlighting the classic bias-variance trade-off. This performance drop is likely due to multicollinearity and redundancy among engineered features, which XGBoost can be particularly sensitive to due to its feature-splitting mechanism mentioned earlier. Additionally, generating a 23-feature input space for a full Landsat scene is memory intensive and impractical for operational deployment. This further highlights the importance of incorporating FS to retain only the most relevant features for the task.

\begin{figure*}[t]
    \centering

    \begin{subfigure}{0.9\textwidth}
        \centering
        \includegraphics[width=\textwidth]{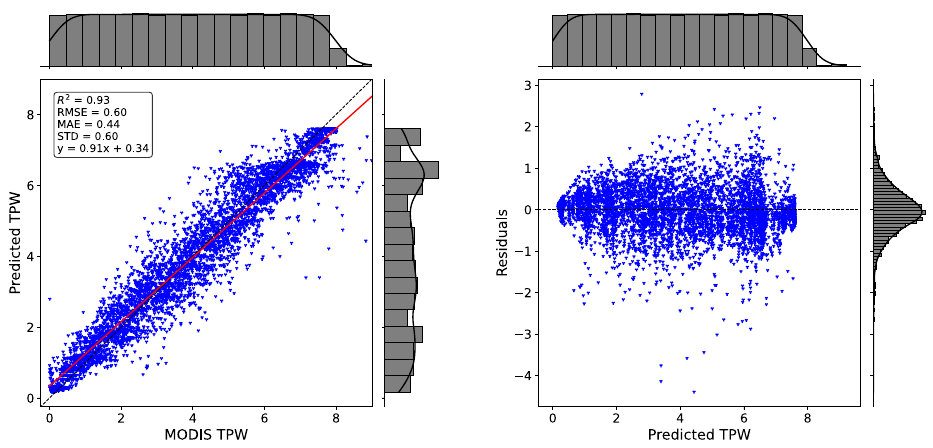}
        \caption*{\textbf{Training Set}}
    \end{subfigure}

    \vspace{1em}

    \begin{subfigure}{0.9\textwidth}
        \centering
        \includegraphics[width=\textwidth]{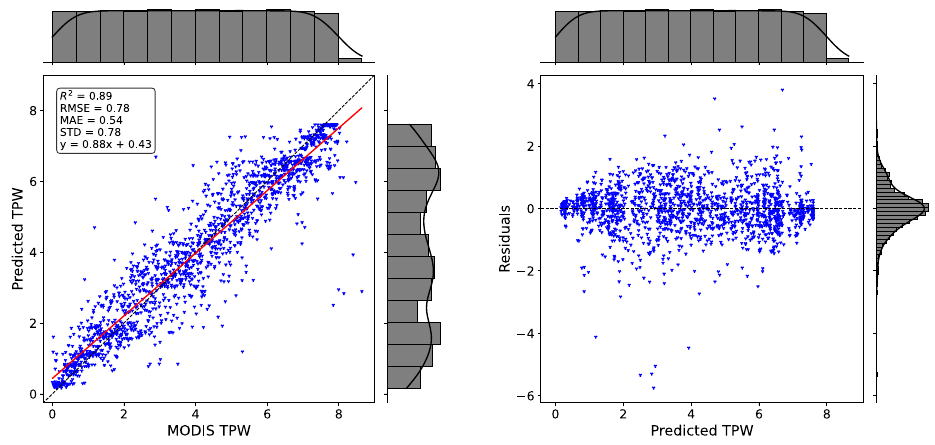}
        \caption*{\textbf{Testing Set}}
    \end{subfigure}

    \caption{\textbf{XGBoost + FE + FS.} Regression plots compare predicted TPW to MODIS TPW, while residual plots show the distribution of prediction errors. Color density represents sample concentration. The workflow accurately captures the nonlinearity in the data.}
    \label{fig:xgboost_regression_residual}
\end{figure*}

We apply both FE and FS on top of XGBoost for improved model performance. The resulting XGBoost + FE + FS model outperforms others, achieving a testing MAE of 0.54 cm and an R² of 0.89 using only three selected features. We present the results for this model in Figure~\ref{fig:xgboost_regression_residual}. Overall, the fitted line closely follows the 1:1 reference line in the regression plot, indicating strong agreement between predicted and actual TPW values. The residual plot shows no apparent trends and appear random. A few outliers with large residuals are visible; these may correspond to edge-of-scene pixels and may require further investigation or marginal data cleaning. 

Moreover, one can observe upper and lower bounds in the predicted values of XGBoost. This is due to the inherent structure of decision trees, which learn limits based on the training data range. While this behavior is expected, it is not a major concern since TPW values above 8 cm are extremely rare in real-world scenarios. 

\begin{figure*}
    \centering
    \includegraphics[width=0.9\linewidth]{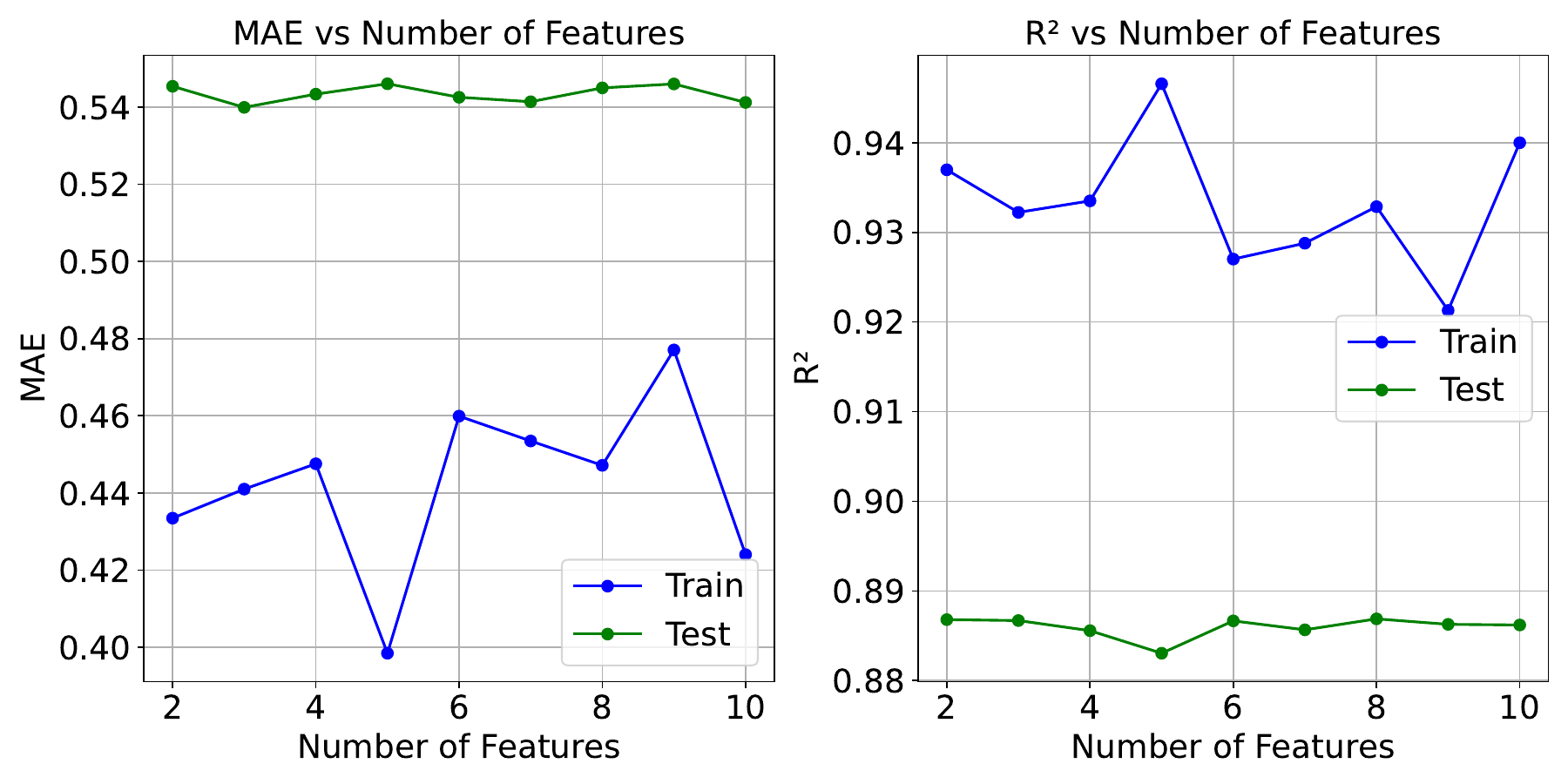}
    \caption{\textbf{XGBoost + FE + FS} vs. number of selected features. We can see marginal improvement in terms of $R^2$ and $MAE$ above 3 features. }
    \label{fig:xgboost_vs_features}
\end{figure*}

In Figure~\ref{fig:xgboost_vs_features}, we show the performance of the XGBoost + FE + FS model as a function of number of features selected. Results indicate that performance plateaus beyond three features, with minimal improvement for $R^2$ and MAE. The three selected features are $\sin(L_{11})$, $(L_{10}/L_{11})^3$, and $\sin(L_{10})$. Notably, the feature $(L_{10}/L_{11})^3$ stands out, as it resembles the MODIS two-band ratio commonly used to infer atmospheric water vapor, potentially capturing similar information between the TIRS bands.

\begin{figure*}
    \centering
    \includegraphics[width=0.8\linewidth]{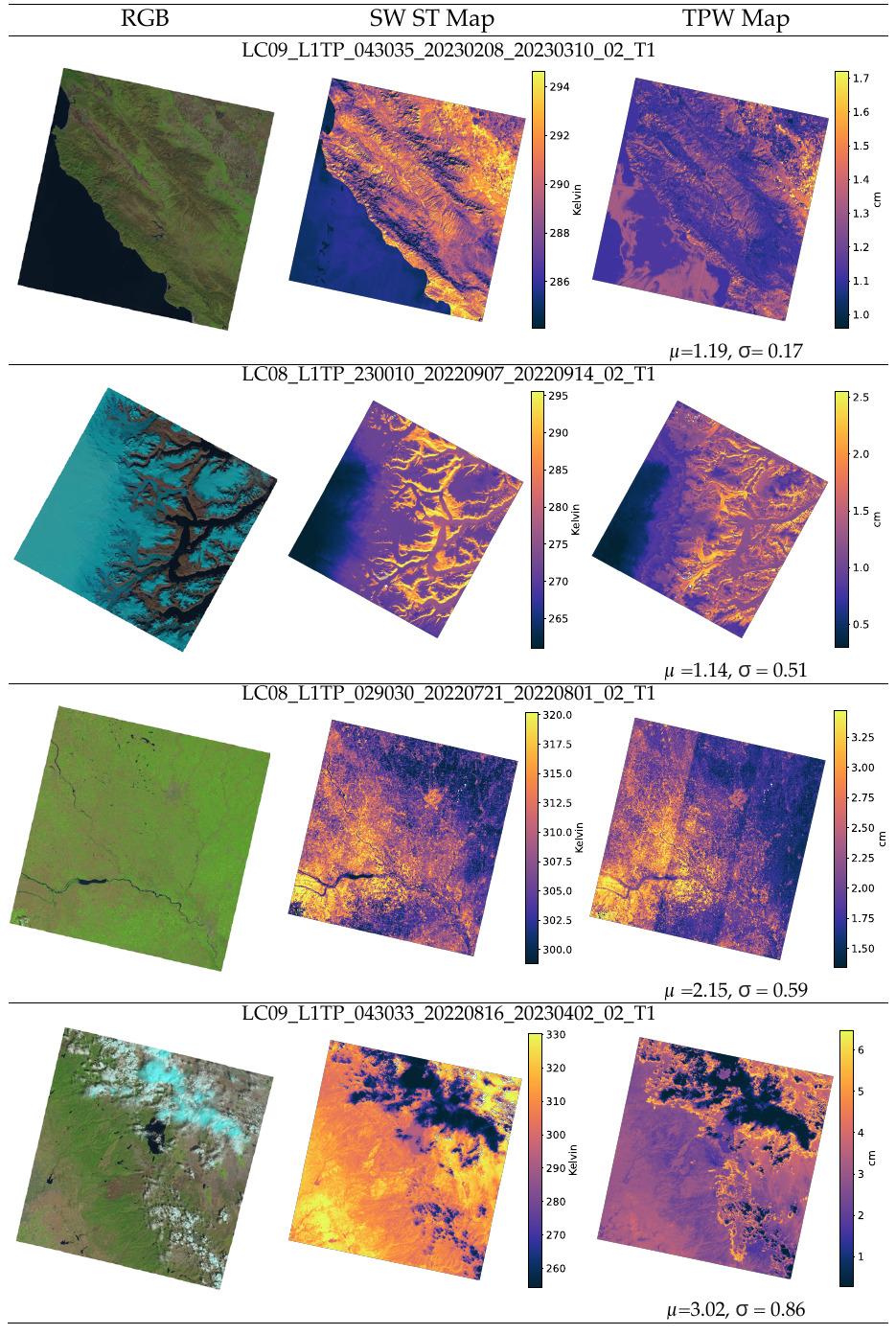}
    \caption{RGB map, SW ST map, and the corresponding estimated TPW map using the proposed algorithm. Note that the colorbars ranges are restricted to  2 and 98 percentiles.}
    \label{fig:scenes_tpw_estimated}
\end{figure*}

We select four Landsat scenes to illustrate generated TPW maps from our proposed algorithm. Figure \ref{fig:scenes_tpw_estimated} shows each scene’s RGB image alongside its corresponding SW ST and TPW maps. We mask out cloud‐covered pixels in TPW map by setting their TPW values explicitly to zero.

We choose the mentioned scenes to represent a diverse range of TPW conditions. The first scene (path/row 43/35) covers Los Angeles, CA in February, where mild temperatures and moderate humidity yield intermediate TPW values. The second scene (path/row 230/10) spans Greenland’s dry, cold environment, producing low TPW values. The third scene, acquired over South Dakota vegetation in summer, exhibits high temperatures and elevated TPW estimates on a cloud-free day. The fourth scene (path/row 43/33) shows that even though we did not train the model on clouds, the model is able to capture the localized high TPW estimates in cloud proximity.

\subsection{Uncertainty Validation}

\begin{figure*}
    \centering
    \includegraphics[width=0.8\linewidth]{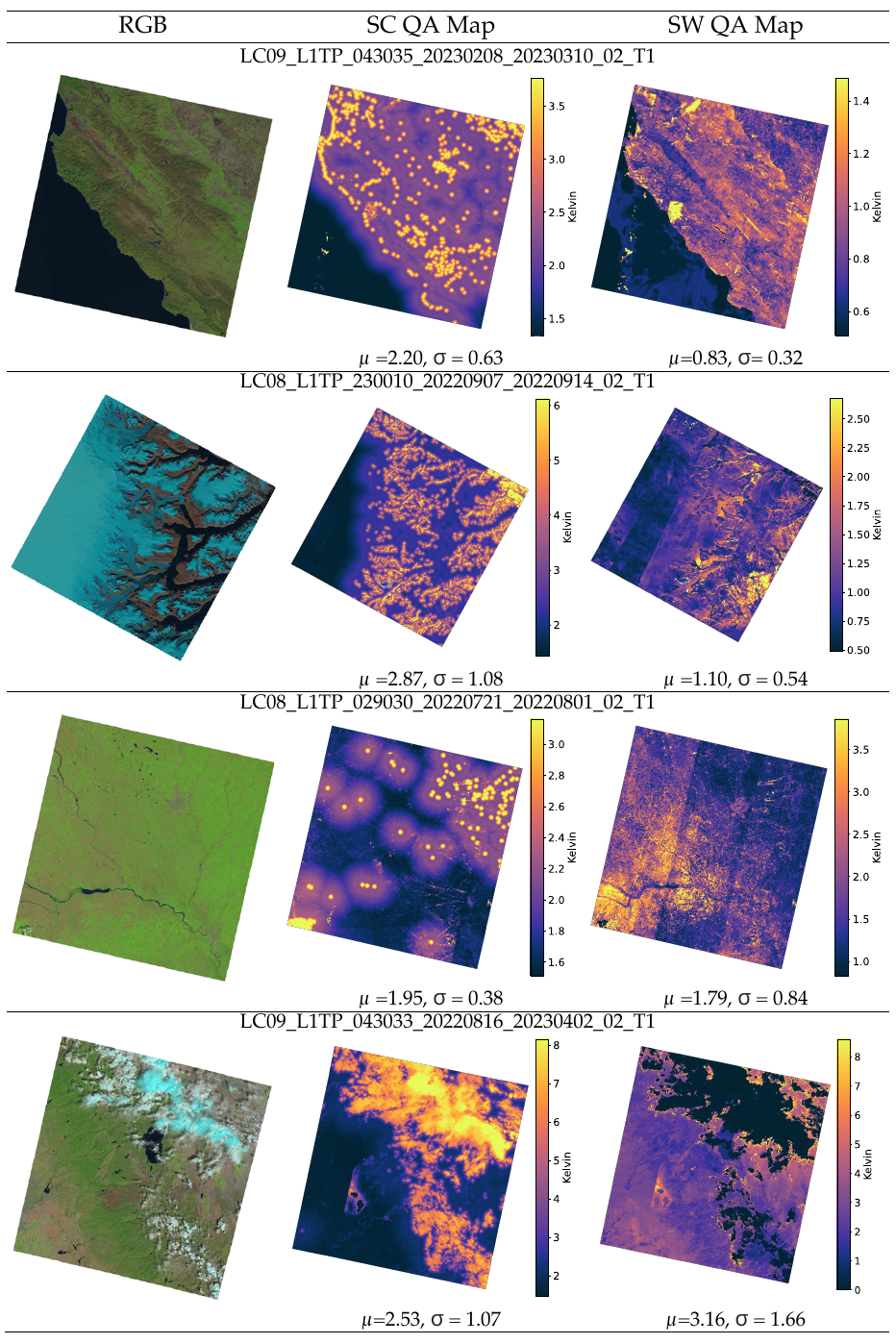}
    \caption{RGB, SC QA and SW QA maps for four different scenes. Note the ``lollipop effect" due to distance-to-cloud parameter in SC-QA map and false positives.}
    \label{fig:uncertainty_samples}
\end{figure*}

In the Methods section, we proposed an uncertainty workflow that produces per-pixel uncertainty maps for the SW approach. Here, we present these uncertainty maps for the four scenes discussed earlier. In Figure~\ref{fig:uncertainty_samples}, we show the RGB image alongside the SC QA map and the SW QA map. As we showed previously, only the fourth scene contains clouds. Note that while the C Function of Mask (CFMask) algorithm,
which populates the cloud confidence mask, is approximately 95\% effective, its susceptibility to false positives (especially on cloud-free days) should prevent it from being used in automated workflows~\cite{foga2017cloud,stillinger2019cloud}. Pixels that appear bright in the reflective with a cold apparent temperature in the thermal will be falsely flagged as clouds (apparent in first three scenes) and, in the case of the current Level 2 workflow, uncertainty is broadly distributed about the affected pixels. On cloud-free days, we can observe false positives over highly reflective, low-emissivity targets such as sand, barns with aluminum roofs, and cities, for example. Note that these scenes are not special cases and an interested user can choose any scene of interest on a
cloud-free day and observe similar anomalies.

Additionally, we compare the mean and standard deviation of uncertainty for the SC and SW workflows. To ensure an unbiased comparison, cloud pixels are masked out using the cloud QA map in both approaches. The resulting plots show that because the SC algorithm relies on the DTC metric, its uncertainty estimates are systematically misleading and inflated for the entire scenes. In the first three scenes (path/row 43/35, 230/10, and 29/30), pronounced ``lollipop” artifacts in the SC QA map drive up its mean and standard deviation. 

Also, considering that Landsat’s current ST retrieval algorithm rely on historical ASTER data for emissivity estimation, all anomalies in ASTER will flow down to the Landsat-based products, see Figure~\ref{fig:uncertainty_samples} (first scene, right column). The
yellow, high-uncertainty, regions highlight these anomalies. Perhaps more egregious to some users is that, due to ASTER’s duty-cycle limitations, only 96.29\%
of worldwide land targets were imaged by ASTER. As such, 3.71\% of the world
will not have an associated Landsat ST-product without gap-filling the affected
areas with an alternative emissivity grid~\cite{astermissing}. We report these percentages based on calculations using Google Earth Engine~\cite{mutanga2019google}.

\subsection{Future Work}

In this study, we leverage a machine learning approach to estimate column water vapor for L8 and L9 using TIRS-class sensors. Deep learning models with convolutional layers are capable of capturing spatial patterns in imagery data that non-convolutional models lack. In theory, the input feature space to a convolutional neural network (CNN) could incorporate the three features identified in this study ($\sin(L_{11})$, $(L_{10}/L_{11})^3$, and $\sin(L_{10})$) serving as substitutes for the RGB channels in encoder–decoder architectures such as U-Net~\cite{ronneberger2015u}. Another key limitation of using CNNs is the need for a sufficiently large and cloud-free training dataset to prevent overfitting, given the increased model complexity. This challenge might become even more significant in high-TPW regions, where persistent cloud cover is common, potentially introducing bias given the imbalanced data.

As for future Landsat satellites and ST models, Landsat Next will be equipped with bands capable of capturing atmospheric water vapor directly (which Landsat 8 and 9 intentionally avoid). This enables collection of richer features across multiple bands. For Landsat Next, a similar workflow proposed in this study, could be adopted that incorporates machine learning along with FE and FS to identify the most contributing features resulting in improved accuracy. We argue that the proposed machine learning framework can also be applied to other products in the Landsat archive, including surface reflectance algorithms such as Sen2Cor and LASRC~\cite{vermote2018lasrc,main2017sen2cor}. For example, Sen2Cor uses a dark object approach for atmospheric correction, estimating aerosol optical thickness (AOT) using a Dense Dark Vegetation (DDV) algorithm that detects dense vegetation, soil, or water bodies. A key limitation of Sen2Cor is reliance on the availability of suitable dark targets in the scene. We introduced a machine learning pipeline that can overcome this limitation by removing the dependency on specific surface types, thereby improving the accuracy and reducing the uncertainty of surface reflectance retrievals.

As for the SW algorithm and its corresponding SW uncertainty approach, a material-specific strategy tailored to different materials categories can further improve accuracy. For example, because water is spectrally flat and uniform, error due to emissivity estimation is lower thus the overall retrieval errors over water are typically much smaller than those over land (of different materials). The difference in the spectral characteristics of various materials suggests that training the SW algorithm with material-specific emissivity profiles can enhance the precision of both the ST estimation and the associated uncertainty quantification.

\section{Conclusions}
The SW algorithm for LST estimation and its various derivations has been extensively studied in the literature for more than three decades~\cite{becker1990towards,wan1996generalized,coll1994atmospheric,ulivieri1994split}. With the launch of L8 and its successor L9 with similar dual long-wave infrared thermal bands onboard, an opportunity for implementation of SW algorithm for ST product presented itself. The current nominal workflow, level-2 SC ST product, is dependent on the DTC metric, resulting in the corresponding SC-QA map being susceptible to overestimation of uncertainty due to false positives originating from CFMask. Starting with Collection 3, the SW algorithm and its corresponding uncertainty map will serve as the level-2 product for ST estimation. This study proposes a SW-specific uncertainty workflow that integrates TPW estimation to improve uncertainty characterization.

Since MODIS and L9 satellites are near-coincidental, this presents an opportunity for TPW estimation using TIRS-2 data. In this study, we suggest a tree-based XGBoost model for TPW estimation based on TIRS-2 bands as independent variables using the MODIS MOD05 product as a reference. Our findings show that TPW estimation based on TIRS-2 bands is feasible with errors as low as $MAE = 0.54 [cm]$ and explained variability as high as $R^2 = 0.89$. 

Moreover, we quantify the uncertainty due to TPW in varying different atmospheres with varying level of TPW using MODTRAN. We observe the dependence and impact TPW has on the estimation of SW-derived ST in Figure \ref{fig:swerror_tpw_relationship}, when compared to reference. The error in estimating ST from the SW-algorithm is correlated with an increase in TPW.

\section*{Contributor Roles Taxonomy}
Amirhossein Hassanzadeh: Methodology, Software, Formal Analysis, Visualization, Investigation, Writing - Original Draft. Robert Mancini: Methodology, Software, Investigation. Aaron Gerace: Supervision, Project administration, Funding acquisition, Conceptualization: Methodology, Validation, Formal Analysis, Visualization, Writing - Original Draft. Rehman Eon: Conceptualization: Methodology, Validation, Software, Formal Analysis, Visualization, Writing - Original Draft. Matthew Montanaro: Conceptualization, Methodology, Writing - Review and Editing.

\section*{Acknowledgements}
This material is based upon work supported by the U.S. Geological Survey under Cooperative Agreement Number G19AC00176. The views and conclusions contained in this document are those of the authors and should not be interpreted as representing the opinions or policies of the U.S. Geological Survey. Mention of trade names or commercial products does not constitute their endorsement by the U.S. Geological Survey. This manuscript is submitted for publication with the understanding that the United States Government is authorized to reproduce and distribute reprints for Governmental purposes.

\bibliographystyle{elsarticle-num-names} 
\bibliography{mybibfile}

\section{Appendix}

\begin{table*}
    \small
    \centering
    \begin{tabular}{p{0.08\textwidth} p{0.35\textwidth} | p{0.08\textwidth} p{0.35\textwidth}}
        ACO & Ant Colony Optimization & AERONET & Aerosol Robotic Network \\
        API & Application Programming Interface &
        ASTER & Advanced Spaceborne Thermal Emission and Reflection Radiometer \\
        ATSR & Along Track Scanning Radiometer &
        AVHRR & Advanced Very High Resolution Radiometer \\
        CAMEL & Combined Aster and MODIS emissivity database over land &
        CFMask & C Function of Mask \\
        DDV & Dense Dark Vegetation &
        DTC & Distance to Cloud \\
        EROS & Earth Resources Observation and Science &
        FE & Feature Engineering \\
        FS & Feature Selection &
        GSD & Ground Sampling Distance \\
        L8 & Landsat 8 &
        L9 & Landsat 9 \\
        LR & Linear Regression &
        LST & Land Surface Temperature \\
        MAE & Mean Absolute Error &
        MLP & Multi-Layer Perceptron \\
        MODIS & Moderate Resolution Imaging Spectrometer &
        MODTRAN & Moderate Resolution Atmospheric Transmission \\
        NDVI & Normalized Difference Vegetation Index &
        NDSI & Normalized Difference Snow Index \\
        NEdT & Noise Equivalent Temperature Difference &
        OLI & Operational Land Imager \\
        QA & Quality Assurance &
        RGB & Red Green Blue \\
        RMSE & Root Mean Square Error &
        RSR & Relative Spectral Response \\
        RTM & Radiative Transfer Model &
        R\textsuperscript{2} & Coefficient of Determination \\
        SC & Single Channel &
        SR & Surface Reflectance \\
        ST & Surface Temperature &
        STD & Standard Deviation \\
        SURFRAD & Surface Radiation Budget Network &
        SW & Split-Window \\
        TES & Temperature Emissivity Separation &
        TIGR & Thermodynamic Initial Guess Retrieval \\
        TIRS & Thermal Infrared Sensors &
        TOA & Top of Atmosphere \\
        TPW & Total Perceptible Water &
        UAS & Unmanned Aerial System \\
        UCSB & University of California, Santa Barbara &
        USGS & United States Geological Survey \\
        VI & Vegetation Index &
        WRS-2 & World Reference System-2 \\
        XGBoost & Extreme Gradient Boosting & & \\
    \end{tabular}
    \caption{Acronyms and abbreviations used in this study.}
    \label{tab:abbreviations}
\end{table*}

\end{document}